\documentclass{aa}  

\usepackage{txfonts}
\usepackage{amsmath,amssymb}
\usepackage{natbib}

\usepackage{hyperref}
\usepackage{url}
\hypersetup{
    colorlinks=true,
    linkcolor=blue,
    citecolor=blue,
    urlcolor=blue
}

\newcommand{\beq}{\begin{equation}}
\newcommand{\eeq}{\end{equation}}
\newcommand{\beqa}{\begin{eqnarray}}
\newcommand{\eeqa}{\end{eqnarray}}

\title{Interstellar Medium Modulation of Nonlinear Kinetic Alfvén Morphology in Structured Galactic Environments}
\titlerunning{ISM Modulation of Nonlinear KA Morphology in Structured Galactic Environments}

\author{
    Manpreet Singh\inst{1,2}\and
    Siming Liu\inst{2}\and
    N. S. Saini\inst{3}
}
\authorrunning{M. Singh et al.}

\institute{
    School of Computing and Artificial Intelligence, Southwest Jiaotong University, Chengdu-610031, PR China.\\
    \email{singhmanpreet185@gmail.com}
    \and
    School of Physical Science and Technology, Southwest Jiaotong University, Chengdu-610031, PR China.
    \and
    Department of Physics, Guru Nanak Dev University, Amritsar-143005, India
}

\date{Received <date>; accepted <date>}

\abstract
{We present a spatially dependent framework for the existence and propagation of nonlinear kinetic Alfvén (KA) structures in the interstellar medium (ISM).}
{By employing a multi-component analytical model, we aim to derive location-dependent coefficients governing KA dispersion and nonlinearity across structured Galactic environments.}
{The model incorporates the diffuse warm ionized medium together with localized H\,II regions, supernova remnants (SNR), and stellar-wind bubbles (SWB). The reductive perturbation method is applied to obtain Korteweg-de Vries (KdV) equations, enabling the characterization of solitons under realistic astrophysical conditions while accounting for superthermality, plasma beta $(\beta)$, temperature, and density gradients.}
{Our results reveal distinct exclusion zones (EZs) for KA solitons in high-$\beta$ H\,II regions and SWB/SNR interiors, as well as ultra low-$\beta$ regions near central pulsar wind nebulae. While H\,II regions exhibit simple Gaussian-driven depletions, the complex ``hole-and-shell'' morphologies of SWBs and SNRs imprint sharp spatial variations and discontinuities on soliton properties.}
{This study establishes a direct link between macroscopic ISM morphology, ion-kinetic scale dissipation, and the emergence of coherent Alfvénic activity. These findings provide a new interpretive framework for radio scattering, pulsar scintillation, and fine-scale signatures in astrophysical observations.}

\keywords{Interstellar medium -- Kinetic Alfvén Waves -- Solitons -- H\,II Regions -- Stellar-Wind Bubbles -- Supernova Remnants.}

\begin{document}

\maketitle

\section{Introduction}

The ionized interstellar medium (ISM) is now firmly established as a turbulent \citep{FERRIEREPlasma2019}, magnetized plasma whose electron density fluctuations span more than ten orders of magnitude in scale. Radio scintillation and scattering measurements reveal a nearly Kolmogorov spectrum extending from $\sim10^{6}$ to at least $\sim10^{13} \mathrm{\,m}$, forming the ``Big Power Law in the Sky’’ \citep{Armstrong1995}. Using H$\alpha$ observations, it was demonstrated \citep{Chepurnov2010} that this turbulent cascade continues to scales approaching several parsecs throughout the warm ionized medium (WIM). Pulsar scintillation studies further show that the cascade reaches down to an inner scale of $10^{5}$–$10^{6}\mathrm{\,m}$ ($\sim100$--$1000\mathrm{\,km}$), comparable to the proton gyroradius \citep{Spangler1990,Rickett2009,OCKERPersistent2021}, implying that interstellar turbulence routinely extends to ion-kinetic scales where fluid MHD breaks down. Interstellar scintillation also reveals intermittent, localized electron density structures with non-Gaussian statistics \citep{BOLDYREVLevy2003,BOLDYREVRadiowave2005,TERRYCoherence2007}, suggesting the presence of steepened, coherent plasma fluctuations that resemble nonlinear dispersive structures.

At sub-ion scales, kinetic theory predicts that the Alfvénic cascade transitions naturally into kinetic Alfvén waves (KAWs), which mediate energy transfer to small scales and ultimately to particle heating and acceleration. Gyrokinetic theory \citep{Schekochihin2009} and solar-wind simulations \citep{Howes2008PRL,Howes2008JGR,Zhao2016} demonstrate that KAWs dominate between ion and electron gyroscales, with dissipation primarily via Landau damping, and recent work shows that KAW intermittency can lead to efficient electron heating \citep{Zhou2023}. Observations indicate that plasma beta in the WIM spans a broad range \citep{Ferriere2001}. While some regions approach $\beta \sim 1$, substantial regions of the diffuse ionized medium exhibit $\beta < 1$, falling within the regime $m_e/m_i \ll \beta \ll 1$ (with $m_e$ and $m_i$ being the electron and ion mass respectively) where KAWs and their nonlinear extensions can exist. Since the WIM is a magnetized, weakly collisional plasma with a cascade extending to ion-kinetic scales, KAW turbulence is a natural candidate for shaping small-scale structure and dissipation in the ISM.

Although the linear and turbulent properties of KAWs are increasingly well understood, the formation of nonlinear coherent KAW structures with steepened fronts, localized wave packets, and solitary pulses has received comparatively little attention in an astrophysical ISM context. Spacecraft measurements in the terrestrial magnetosphere reveal localized, dispersive Alfvén/kinetic-Alfvén wave packets carrying substantial Poynting flux into the auroral region \cite[e.g.][]{CHASTONKinetic2003,CHASTONTurbulent2008}, and the review by \citet{Stasiewicz2000} summarizes extensive observations and theory of small-scale Alfvénic structures, including density cavities, vortices, and steepened fronts. 
Analytical and observational studies further show that finite-amplitude kinetic/inertial Alfvén waves can undergo nonlinear steepening and form localized,
coherent structures under suitable plasma conditions \citep{Stasiewicz2000,SinghKourakis2019}. However, these works focus on localized, relatively homogeneous space or planetary plasma environments and do not address how coherent KAW structures develop within the highly structured Galactic ISM. Intermittent small-scale density and magnetic structures inferred in the ionized ISM indicate that coherent nonlinear features may contribute to dissipation, consistent with the broader picture from turbulence studies in which intermittent, front-like structures can dominate energy transfer and dissipation \citep{Warhaft2000}.

The ISM is highly non-uniform, comprising structures such as H\,II regions, supernova remnants (SNRs), stellar-wind bubbles (SWBs) etc., each producing sharp gradients in density, temperature, magnetic field strength, and plasma beta \citep{Ferriere2001,Elmegreen2004,MacLow2004}. Stellar feedback injects energy on parsec scales through supernova explosions, expanding H\,II regions, and SWBs, maintaining a mildly supersonic, magnetized turbulent cascade across the WIM \citep{Hill2012}. Furthermore, the interiors of SNRs, particularly near Pulsar Wind Nebulae (PWNe), are enriched with relativistic pairs, creating a multi-species electron-positron-ion (although with trace ions) plasma that fundamentally alters wave dispersion.
Modern numerical studies indicate that the ISM often operates in a strong-turbulence regime where magnetic fluctuations reach $|\delta\mathbf{B}| \gtrsim |\mathbf{B}_0|$ \citep{Goldreich1995,Perez2008,Schekochihin2022}, and galaxy-scale simulations show that such large-amplitude disturbances are ubiquitous and driven by stellar feedback \citep{Ntormousi2024}. Strong turbulence naturally generates intermittent coherent structures, such as current sheets, filaments, vortices, and shocklets, that localize dissipation and promote nonlinear wavefront steepening \citep{Uritsky2010}. Such structures are well known to excite finite-amplitude Alfvénic and kinetic-Alfvénic fluctuations \citep{KARIMABADICoherent2013}, providing physically motivated seed perturbations that can steepen into localized kinetic Alfvén (KA) solitons under appropriate plasma conditions \citep{SinghKourakis2019}. While current sheets, shocklets, and magnetic filaments are well-established products of strong turbulence, the possibility that the same turbulent dynamics can generate KA solitons has not yet been explored in the context of the ISM. If present, KA solitons can produce localized density and magnetic-field fluctuations that influence observables such as radio-wave scattering, pulsar scintillation, or microstructures in timing data, providing potential astrophysical signatures of coherent KA activity.

Because the ISM contains strong spatial variations in magnetic field, density, temperature, and particle composition, the coefficients governing KA dispersion and nonlinearity can vary significantly from region to region. Even modest environmental changes can modify soliton amplitude, width, or existence conditions. 
Thus, the structured nature of the ISM, including high-$\beta$ H\,II cavities, compressed SNR shells, and density-evacuated SWBs can directly influence the formation, propagation, and characteristics of nonlinear KA solitons \citep{Subramanian2006,Kritsuk2017}. In addition to their spatial variability, fine-scale KA structures are efficient sites of electron heating in weakly collisional plasmas \citep{TenBarge2013}, indicating that the soliton structures considered here may influence local ISM thermodynamics.

The velocity distribution of electrons in the ISM can deviate from a strict Maxwellian, particularly in weakly collisional plasmas subject to continuous driving and intermittent energization. These conditions produce suprathermal power-law tails, which are often modeled using the $\kappa$-distribution, a framework originally developed for space plasma contexts and now standard for characterizing non-thermal populations \cite[e.g.,][]{VASYLIUNASSurvey1968, SUMMERSModified1991, PIERRARDKappa2010, LIVADIOTISKappa2009}. In this formulation, the $\kappa$-index quantifies the departure from thermal equilibrium: a smaller $\kappa$ corresponds to a harder high-energy tail, while the Maxwellian distribution is smoothly recovered as $\kappa \rightarrow \infty$ (in practice, $\kappa \gtrsim 100$ is often indistinguishable from a Maxwellian).
Within ISM context, $\kappa$-distributed electrons have been proposed for photoionized nebulae (H\,II regions and planetary nebulae), although this specific application has been challenged by work arguing that the electron energy distribution is extremely close to Maxwellian under typical nebular conditions \citep{NICHOLLSResolving2012c,DRAINEElectron2018a}. 
More broadly, in the structured ISM, persistent suprathermal electrons are physically expected where continuous energization competes with thermalization. This includes supernova-driven collisionless shocks and superbubbles, which generate non-thermal tails via diffusive shock acceleration and collective acceleration mechanisms, respectively, as well as turbulence enabled magnetic reconnection, which can efficiently accelerate particles and produce nonthermal spectra \citep{DRURYIntroduction1983,PARIZOTSuperbubbles2004,LAZARIANReconnection1999,GUOFormation2014a}. 
In our KA soliton context, adopting $\kappa$ is therefore not merely phenomenological: it changes the electron kinetic response that closes the parallel electric field and charge-separation physics underlying dispersive KAW dynamics, and thereby provides a physically motivated control on KA soliton morphology most strongly on the soliton width, especially near sharp gradients and kinetic regime boundaries \citep{MASOODNonlinear2015}

In this work, we develop the first spatially dependent framework for nonlinear KA solitons in a realistic Galactic ISM. We construct a multi-component analytical model incorporating WIM background profiles together with embedded H\,II regions, SWBs, and SNRs, and use the resulting spatial distributions of plasma parameters to derive location dependent KdV coefficients governing KA dispersion and nonlinearity, explicitly accounting for the kinetic effects of suprathermal electrons. This enables the first Galactic maps of KA soliton existence, amplitude, and width. We identify distinct KA soliton exclusion zones linked to high-$\beta$ H\,II regions and ultra low-$\beta$ SNR interiors, and show how ISM structures imprint sharp spatial variations on soliton properties. These results establish a direct connection between macroscopic ISM structure, turbulent energy dissipation at ion kinetic scales, and the formation of nonlinear KA solitons in astrophysical plasmas.

\section{Theoretical Model \label{sec:Theoretical_Model}}

The ISM is a vast, multi-phase plasma environment governed by the gravitational and magnetic fields of the Milky Way. Its inherent inhomogeneity leads to the modulation of the properties of any propagating wave by its location and proximity to different galactic structures, including H\,II regions, SWBs, and SNRs, among others.
To model this, we distinguish between the global coordinate system of the galaxy and the local coordinate system in which the wave dynamics are resolved.

\medskip
\noindent
\textbf{Global Coordinates:} The large scale background plasma parameters are described in a cylindrical galactic coordinate system $(R, {Z})$, where $R$ is the galactocentric radius and ${Z}$ is the vertical height from the galactic mid-plane.

\medskip
\noindent
\textbf{Local Coordinates:} The wave dynamics are analyzed in a local Cartesian coordinate system $(x, y, z)$.
We make the standard approximation that, at any specific galactic location $(R, Z)$, the large-scale Galactic magnetic field $\mathbf{B}$ can be treated as locally uniform and aligned with the $z$-axis of this frame, allowing the KA solitons to propagate obliquely in the $x$-$z$ plane. 
This alignment is a common assumption in case of obliquely propagating dispersive Alfvén waves, as the ISM's complex magnetic structure, predominantly azimuthal (spiral) in the disk with pitch angles of $\sim 10$--$30^\circ$ and poloidal/toroidal components in the halo, varies slowly over the small scales of wave propagation, where the characteristic KA wavelength (comparable to the ion gyroradius, $\rho_i \sim 200$ km) is many orders of magnitude smaller than the kiloparsec-scale background gradients \citep{Jansson2012}.
By rotating the local frame to match the instantaneous $\mathbf{B}$ direction at $(R, Z)$, we capture the essential oblique propagation of KA solitons, characterized by $k_\perp \gg k_\parallel$ (where $k_\perp$ and $k_\parallel$ are perpendicular and parallel wave vectors respectively; \citet{Hollweg1999}).
This approximation is validated in gyrokinetic simulations, where field curvature effects are negligible for sub-ion scales \citep{Schekochihin2009}.
This scale separation allows us to link micro-scale plasma physics with macro-scale astrophysical structures, determining soliton properties based on their specific galactic location $(R,Z)$ while treating the local background as homogeneous.

\subsection{Multi-Component Representation of the ISM \label{subsec:multi-component}}

The real ISM is not homogeneous but instead composed of a smooth diffuse background with superposed localized structures created by stellar evolution. To capture this clumpy, multi-phase environment in a self-consistent manner, we adopt the principle of linear superposition, in which every physical quantity $Q$ (e.g., magnetic field, plasma density, and temperature) at any given position is written as the sum of a background component and the localized contributions from all embedded structures:
\begin{equation}
Q_{\mathrm{total}}(R,Z) = Q_{\mathrm{bkg}}(R,Z) + \sum_{i} Q_{\mathrm{struct}, i}(R,Z).
\label{eq:superposition}
\end{equation}
For each structure, $Q$ is modeled using Gaussian, super-Gaussian or shell-type analytical forms, selected for consistency with observations and ease of integration into the plasma model.

Below, we describe the background ISM and each embedded structure individually, specifying for each its electron density, magnetic field, temperature, and positron fraction (where applicable). 
We parameterize the model in terms of electron quantities (e.g., density $n_{e,\text{total}}$ and temperature $T_{e,\text{total}}$) because these are the primary observables constrained by astronomical surveys (e.g., pulsar dispersion measures, thermal emission lines). Complementing this, we incorporate a modeled positron fraction ($p_{\text{total}}$) based on standard PWN theories to accurately represent pair-enriched environments where $n_{e,\text{total}}$ significantly exceeds ion density $n_{i,\text{total}}$. However, since KAWs are fundamentally ion-scale modes governed by ion inertia and finite Larmor radius effects, the corresponding ion density $n_{i,\text{total}}$, which sets the local Alfvén speed, is subsequently derived from these electron and positron profiles via charge neutrality.

\subsection{Analytical Formalism for Localized Structures}
\label{subsec:Analytical_formalism}
To ensure a consistent mathematical description across the various ISM structures (H\,II regions, SWBs, and SNRs) and avoid redundancy, we utilize three generalized analytical functions to model the spatial profiles of density, magnetic field, and temperature. The spatial variation is parameterized by the radial distance $d_i$ from the center $(R_i, Z_i)$ of the $i$-th structure:
\begin{equation}
d_i = \sqrt{(R - R_i)^2 + (Z - Z_i)^2}.
\end{equation}

\noindent
\emph{Gaussian Profile:} Localized enhancements (e.g., H\,II regions, PWN cores) or depletions (e.g., cavities) are modeled using a standard Gaussian function $\mathcal{G}$:
\begin{equation}
\mathcal{G}(d; A, \sigma) \equiv A \exp\left[-\frac{d^2}{2\sigma^2}\right],
\label{eq:generic_gaussian}
\end{equation}
where $A$ is the amplitude (positive for enhancement, negative for depletion) and $\sigma$ is the characteristic spatial width.

\noindent
\emph{Super-Gaussian Profile:} To represent isobaric regions with flat-top profiles (e.g., SWB interiors), we define a super-Gaussian function $\mathcal{P}$ (Plateau):
\begin{equation}
\mathcal{P}(d; A, R_{\mathrm{flat}}, s) \equiv A \exp\left[-\left(\frac{d}{R_{\mathrm{flat}}}\right)^s\right],
\label{eq:generic_supergauss}
\end{equation}
where $R_{\mathrm{flat}}$ determines the extent of the plateau and the exponent $s$ (typically $s=8$) controls the steepness of the cutoff.

\noindent
\emph{Asymmetric Shell Profile:} To capture the shock-compressed morphology of SNRs and SWBs, we define a piecewise asymmetric shell function $\mathcal{S}$. This function peaks at the shell radius $R_{\mathrm{sh}}$ and decays with different characteristic length scales inward and outward:
\begin{multline}
\mathcal{S}(d; A, R_{\mathrm{sh}}, \sigma_{\mathrm{in}}, \sigma_{\mathrm{out}}) 
\equiv
A \begin{cases}
\exp\left[-\dfrac{(R_{\mathrm{sh}} - d)^2}{2\sigma_{\mathrm{in}}^2}\right], & d \le R_{\mathrm{sh}}, \\[10pt]
\exp\left[-\dfrac{(d - R_{\mathrm{sh}})^2}{2\sigma_{\mathrm{out}}^2}\right], & d > R_{\mathrm{sh}},
\end{cases}
\label{eq:generic_shell}
\end{multline}
where $A$ is the peak amplitude at the shell, and $\sigma_{\mathrm{in}}$ and $\sigma_{\mathrm{out}}$ govern the gradients of the interior shoulder and the exterior shock jump, respectively.

\subsection{Modeling of ISM Components}
\label{subsec:Physical_Modeling_of_ISM_Components}
Now, we construct a spatially dependent model of the ionized ISM that can be used to evaluate the local plasma parameters entering the KA-soliton framework. The large-scale WIM is modeled as a smooth Galactic background, while discrete astrophysical structures are treated as localized perturbations superposed on this background using the analytical profiles introduced in Sec. 
\ref{subsec:Analytical_formalism}.

\noindent
\subsubsection{Galactic Background}
\label{subsubsec:Galactic_background}

The background electron density, $n_{e,\mathrm{bkg}}$, represents the diffuse WIM. It is modeled after the NE2001 framework \citep{cordes2002ne2001}, with typical densities of $\sim 0.01$--$0.1 \,\text{cm}^{-3}$ \citep{GAENSLERVertical2008a} sustained by the leakage of far-ultraviolet radiation from the galactic disk \citep{Haffner2009}. The density profile is defined as:
\begin{equation}
n_{e,\mathrm{bkg}}(R,Z) = n_1 e^{-|Z|/h_1} + n_2 e^{-|Z|/h_2 - (R-R_\odot)/L_2},
\end{equation}
where $n_{1}$ and $h_{1}$ are the number density and vertical scale height of the thin galactic disk; $n_{2}$, $h_{2}$, and $L_{2}$ are the number density, vertical scale height, and radial scale length of the thicker WIM disk component; and $R_{\odot}$ is the galactocentric radius of the Sun (typically 8.5 kpc).


Unlike electrons and ions, which constitute the primordial thermal plasma, positrons are secondary particles produced by energetic processes \citep{WEIDENSPOINTNER2008,PRANTZOS2011}. While they eventually annihilate, a diffuse population is maintained in dynamic equilibrium by continuous injection from cosmic-ray spallation \citep{MOSKALENKOProduction1998} and radioactive decay \citep{KNODLSEDERAllsky2005}. We model the total positron fraction as the ratio of local positron density, $n_{p,\text{total}}$ to the local ion density, $n_{i,\text{total}}$ i.e., $p_{\text{total}} \equiv n_{p,\text{total}} / n_{i,\text{total}}$.
This is constructed as a linear superposition of a large-scale Galactic background and localized sources. The large-scale component, $p_{\text{large-scale}}(R)$, consists of a small, uniform background, $p_{\text{bkg}}$, representing the steady-state diffuse population, and a centrally concentrated Galactic source, $p_{\text{GC}}(R)$:
\begin{equation}
p_{\text{large-scale}}(R) = p_{\text{bkg}} + p_{\text{GC}}(R).
\end{equation}
The Galactic center source is modeled as a Gaussian peak centered at $R=0$. To ensure the total fraction at the center equals the maximum observed value $p_{\text{peak}}$, the source amplitude is defined as the excess above the background:
\begin{equation}
p_{\text{GC}}(R) = (p_{\text{peak}} - p_{\text{bkg}}) \exp\left[-\frac{R^2}{2\,\sigma_{p,\text{scale}}}\right], 
\end{equation}
where $\sigma_{p,\text{scale}}$ is the Galactic positron source scale. H\,II regions and SWBs contribute negligibly to pair production; therefore, we do not include separate contributions from these sources in the total positron fraction. The dominant \emph{local} source of positrons relevant for our model comes from SNR and their associated pulsars, which we describe in the section on SNR.

The Galactic magnetic field is a crucial agent in ISM dynamics, and its structure is similarly multi-scale. The total magnetic field strength is written as a sum of large-scale background component and self-consistent local perturbations from astrophysical structures. The background field, $B_{0,\text{bkg}}$, is based on the disk-halo paradigm, consistent with galactic dynamo theory and extensive observational data from Faraday rotation and synchrotron emission \citep{Jansson2012}. This component represents the large-scale, ordered field generated by the differential rotation of the galactic disk. So the large-scale galactic magnetic field can be written as:
\begin{multline}
B_{0,\mathrm{bkg}}(R,Z) =
B_\odot e^{-(R-R_\odot)/L_B}e^{-|Z|/h_{B,1}} 
+\frac{B_{\mathrm{halo}}}{1+(|Z|/h_{B,2})^2},
\end{multline}
where $B_\odot$ is the reference magnetic field strength at the Sun's galactocentric radius ($R_\odot$); $L_B$ and $h_{B,1}$ are the radial and vertical scale lengths for the disk component, respectively; $B_{\text{halo}}$ is the characteristic field strength in the galactic halo, and $h_{B,2}$ is its vertical scale height.

The ISM exists in a state of thermal multiphase equilibrium, where regional temperatures vary significantly depending on the local balance of heating and cooling mechanisms \citep{Ferriere2001}. Our model for the total plasma temperature, $T_{e,\mathrm{total}}$, reflects this by superposing intense local heating sources onto a cooler, quasi-isothermal background. The background temperature, $T_{e,\mathrm{bkg}}$, is set to the characteristic value of the WIM: $T_{e,\mathrm{bkg}} \approx 8000\,\mathrm{K}$.
This temperature is maintained by a delicate equilibrium between large-scale, diffuse heating from the Galaxy's far-ultraviolet radiation field \citep{Haffner2009} and cooling via atomic forbidden-line emission from ions such as O$^{+}$ and N$^{+}$ \citep{OSTERBROCK2006a}.

\subsubsection{Model of H II Regions}
\label{subsubsec:HII_Region}

A detailed analytical formulation of the H\,II region model is provided in Appendix \ref{subsubsec:HII_Region}.

\subsubsection{Model of Stellar-Wind Bubbles}
\label{subsubsec:SWB}
The full analytical development of the SWB model is presented in Appendix \ref{subsubsec:SWB}.

\subsubsection{Model of Supernova Remnants}
\label{subsubsec:SNR}

The expanding forward shock from an SNR sweeps up the surrounding interstellar gas, compressing the plasma into a dense, turbulent shell while simultaneously evacuating the interior \citep{CHEVALIERInteraction1977,VINKSupernova2012}. This process results in a complex density morphology consisting of a swept-up shell and a depleted cavity. In composite SNRs, this cavity also hosts a central PWN, injected by the relativistic wind of the compact remnant \citep{KENNELConfinement1984a,GAENSLEREvolution2006a}.
The shell is characterized by a strong peak in electron density. This enhancement is highly asymmetric, exhibiting a sharp density jump at the outer shock front followed by a gradual decline toward the interior due to post-shock relaxation \citep{REYNOLDSModels1998,ORLANDOOrigin2007}. The blast wave efficiently clears ambient ions from the remnant's interior, creating a deep density cavity. Additionally, the PWN acts as an efficient source for electron-positron pairs, locally elevating the electron density in the core.
Using the formalism from 
Sec.~\ref{subsec:Analytical_formalism},
we model this three-component morphology as a superposition of the PWN core, the cavity, and the shell:
\begin{multline}
n_{e,\mathrm{SNR}}(R,Z) = 
\mathcal{G}(d_{\mathrm{SNR}}; A_{n,\mathrm{PWN,SNR}}, \sigma_{n,\mathrm{PWN,SNR}}) \\
+ \mathcal{G}(d_{\mathrm{SNR}}; A_{n,\mathrm{cav,SNR}}, \sigma_{n,\mathrm{cav,SNR}}) \\
+ \mathcal{S}(d_{\mathrm{SNR}}; A_{n,\mathrm{sh,SNR}}, R_{\mathrm{sh,SNR}}, \sigma_{n,\mathrm{in,SNR}}, \sigma_{n,\mathrm{out,SNR}}),
\label{eq:SNR_density_total}
\end{multline}
where $d_{\mathrm{SNR}} = \sqrt{(R - R_{3})^2 + (Z - Z_{3})^2}$ is the radial distance from the SNR center $(R_3, Z_3)$. Here, $A_{n,\mathrm{PWN,SNR}}$ is the electron density amplitude of the central pair plasma and $A_{n,\mathrm{cav,SNR}}$ is the negative amplitude representing the interior depletion. For the shell component, $A_{n,\mathrm{sh,SNR}}$ and $R_{\mathrm{sh,SNR}}$ denote the peak density and shell radius, while $\sigma_{n,\mathrm{in,SNR}}$ and $\sigma_{n,\mathrm{out,SNR}}$ govern the widths of the interior shoulder and the exterior shock jump, respectively. This composite model accurately reproduces the island and shell morphology observed in composite SNRs.

Since the PWN is a source of relativistic pairs, the positron density follows the same profile as electrons in the PWN \citep{KENNELConfinement1984a,GAENSLEREvolution2006a}. We model this central positron source as a Gaussian distribution:
\begin{equation}
p_{\mathrm{SNR}}(R,Z) = \mathcal{G}(d_{\mathrm{SNR}}; A_{p,\mathrm{SNR}}, \sigma_{p,\mathrm{SNR}}),
\end{equation}
where $A_{p,\mathrm{SNR}}$ is the peak positron fraction. The spatial extent $\sigma_{p,\mathrm{SNR}}$ is set equal to the electron density width of the PWN (i.e., $\sigma_{p,\mathrm{SNR}} = \sigma_{n,\mathrm{PWN,SNR}}$), indicating that pair production is coincident with the PWN core.

In our SNR model, the magnetic field is structured in close analogy to the plasma density. The same forward shock that sweeps up and compresses the gas also compresses and amplifies the magnetic field lines via flux freezing \citep{VINKSupernova2012}. Consequently, the magnetic morphology exhibits a turbulent, high-field shell surrounding a comparatively weaker interior cavity that separates the shell from the central PWN core  \citep{KENNELConfinement1984a,GAENSLEREvolution2006a}.
We model the total SNR magnetic field as the superposition of the shell and the PWN core:
\begin{multline}
B_{0,\mathrm{SNR}}(R,Z) = 
\mathcal{G}(d_{\mathrm{SNR}}; A_{B,\mathrm{PWN,SNR}}, \sigma_{B,\mathrm{PWN,SNR}})\\
+\mathcal{S}(d_{\mathrm{SNR}}; A_{B,\mathrm{sh,SNR}}, R_{\mathrm{sh,SNR}}, \sigma_{B,\mathrm{in,SNR}}, \sigma_{B,\mathrm{out,SNR}})
\label{eq:SNR_B_total}
\end{multline}
Here, the central Gaussian term represents the PWN, where $A_{B,\mathrm{PWN,SNR}}$ and $\sigma_{B,\mathrm{PWN,SNR}}$ denote the core field strength and its spatial extent, respectively. The shell component is defined by the peak amplitude $A_{B,\mathrm{sh,SNR}}$ and radius $R_{\mathrm{sh,SNR}}$, while $\sigma_{B,\mathrm{in,SNR}}$ and $\sigma_{B,\mathrm{out,SNR}}$ govern the interior and exterior magnetic gradients. 
The condition $\sigma_{B,\mathrm{PWN,SNR}} \ll R_{\mathrm{sh,SNR}}$ ensures that the PWN field dominates only in the innermost region, while the shell term captures the shock amplification.

While the plasma density and magnetic field exhibit a hollow cavity structure, the temperature profile follows a distinctly different morphology. At the outer shock front, the kinetic energy of the sweeping ejecta is thermalized, creating an immediate jump in temperature \citep{CHEVALIERInteraction1977,VINKSupernova2012}. Interior to the shell, the physics of the Sedov-Taylor expansion phase dictates that while the density drops, the pressure remains high to support the expanding shell \citep{CHEVALIERInteraction1977}. Consequently, the temperature must rise towards the center ($T \propto P/n$) to maintain an approximately isobaric state. Furthermore, the center of the remnant is energized by the relativistic PWN \citep{GAENSLEREvolution2006a,KENNELConfinement1984a}.
To model this rising interior thermal structure, we treat the temperature as a superposition of shock heating at the shell and a broad, hot core filling the interior:
\begin{multline}
T_{e,\mathrm{SNR}}(R,Z) = 
\mathcal{G}(d_{\mathrm{SNR}}; A_{T,\mathrm{core,SNR}}, \sigma_{T,\mathrm{core,SNR}}) \\
+ \mathcal{S}(d_{\mathrm{SNR}}; A_{T,\mathrm{sh,SNR}}, R_{\mathrm{sh,SNR}}, \sigma_{T,\mathrm{in,SNR}}, \sigma_{T,\mathrm{out,SNR}}).
\end{multline}
Here, $A_{T,\mathrm{core,SNR}}$ represents the peak temperature of the hot interior bubble. For the shell component, $A_{T,\mathrm{sh,SNR}}$ denotes the temperature excess at the shell radius $R_{\mathrm{sh,SNR}}$, while $\sigma_{T,\mathrm{in,SNR}}$ and $\sigma_{T,\mathrm{out,SNR}}$ characterize the thermal gradients on the interior and exterior sides, respectively. The core width is set broadly ($\sigma_{T,\mathrm{core,SNR}} \approx R_{\mathrm{sh,SNR}}/5$) to ensure the high temperature fills the cavity. This model accurately reproduces the thermal profile observed in composite SNRs, where the tenuous cavity remains the hottest region of the nebula \citep{TEMIMLATETIME2015}.

\subsection{Total ISM Profiles and Equilibrium Ion Density}
\label{subsec:total_ISM_Ion_Density}

The complete astrophysical environment for our soliton analysis is obtained by summing the background and structural contributions according to the principle of linear superposition. Note that since the individual structure functions 
explicitly account for depletions via negative amplitudes, the total profiles are simple summations:
\begin{align}
n_{e,\mathrm{total}}(R,Z)
&= n_{e,\mathrm{bkg}}
 + n_{e,\mathrm{HII}}
 + n_{e,\mathrm{SNR}}
 + n_{e,\mathrm{SWB}}, \\
B_{0,\mathrm{total}}(R,Z)
&= B_{0,\mathrm{bkg}}
 + B_{0,\mathrm{HII}}
 + B_{0,\mathrm{SNR}}
 + B_{0,\mathrm{SWB}}, \\
T_{e,\mathrm{total}}(R,Z)
&= T_{\mathrm{e,bkg}}
 + T_{\mathrm{e,HII}}
 + T_{e,\mathrm{SNR}}
 + T_{e,\mathrm{SWB}}, \\
p_{\mathrm{total}}(R,Z)
&= p_{\text{large-scale}}(R)
 + p_{\mathrm{SNR}}\label{p-total}.
\end{align}

\noindent
While the NE2001 \citep{cordes2002ne2001} and JF12 \citep{Jansson2012} models provide robust large-scale baselines, our adoption of Gaussian and shell-like parameterizations offers a mathematically tractable approach for representing localized plasma morphologies. This framework captures the dominant gradient variances essential for wave modulation, though it may smooth over fine-scale turbulent clumpiness observed in high-resolution maps \citep{LEROYCLUMPING2013,Yao2017}.

A critical step in linking our astrophysical ISM model to the plasma fluid equations is to determine the local equilibrium ion number density, $n_{i,\text{total}}(R,Z)$, which is the fundamental parameter for ion-scale wave phenomena. Our ISM model directly provides the total electron number density and the positron fraction. These quantities are linked to the ion density through the charge neutrality condition:
\begin{equation}
n_{e,\text{total}}(R,Z) = n_{i,\text{total}}(R,Z)+n_{p,\text{total}}(R,Z).
\end{equation}
Using our definition of the positron fraction, $p_{\text{total}}$, we can express the equilibrium ion number density entirely in terms of our model's primary outputs:
\begin{equation}
n_{i,\text{total}}(R,Z) = \frac{n_{e,\text{total}}(R,Z)}{1+p_{\text{total}}(R,Z)}.
\label{ion_density_derivation}
\end{equation}

\noindent
We emphasize that while $p_{\text{total}}$ is negligible in H\,II regions and SWBs, it becomes the dominant factor in the interiors of composite SNRs, where the injection of relativistic pairs leads to significant ion depletion. This derived density $n_{i,\text{total}}(R,Z)$ serves as the fundamental reference for determining the local Alfvén speed and normalizing the fluid equations in the subsequent analysis.

\begin{figure}[!h]
\centering
\includegraphics[width=0.47\textwidth,height=0.705\textwidth]{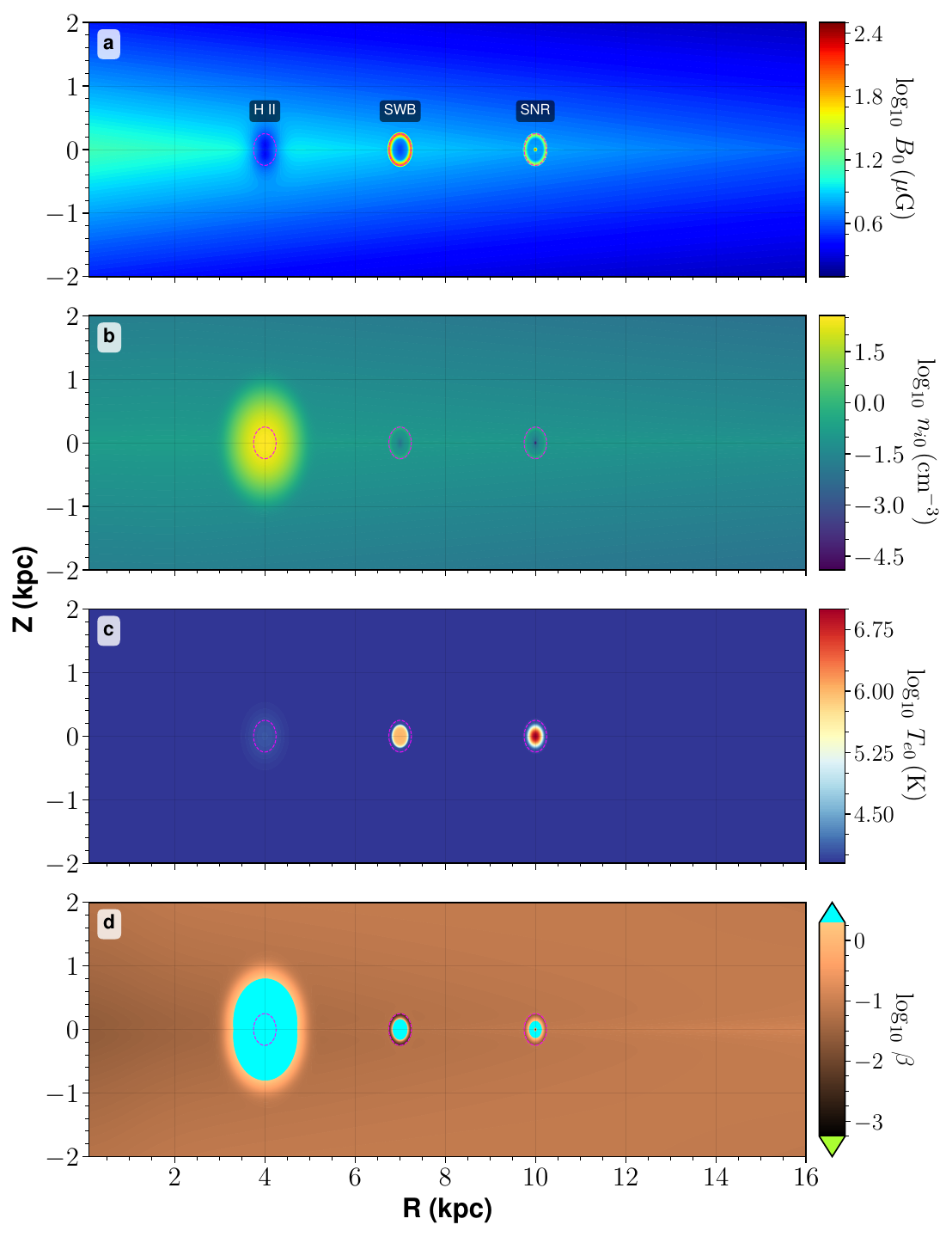}
\caption{Contour plots of key ISM parameters in the local galactic disk plane near embedded astronomical structures. Radius of magenta dashed circles represent the characteristic size of the embedded structures.  
Panel (a) shows $\log_{10}$ of total magnetic field strength $B_0$ \textbf{($\mu$G)}, featuring diamagnetic depletion in the H\,II region and shell like compressions at the SWB and SNR.
Panel (b) shows $\log_{10}$ of total ion number density $n_{i0}$ (cm$^{-3}$).
Panel (c) shows $\log_{10}$ of electron temperature $T_{e0}$ (K).
Panel (d) shows $\log_{10}$ of plasma $\beta$;
high-$\beta$ ($\beta > 1$) regions are masked in cyan, while ultra low-$\beta$ ($\beta < m_e/m_i$) regions are masked in green.
The H\,II region is centered at $(R, Z) \approx (4,0)\,$kpc, SWB at $(7,0)\,$kpc and SNR at $(10,0)\,$kpc. Note that the background field in panel (a) is normalized to $\mu$G to match the 1D profiles in Figs.~\ref{fig: swb} and \ref{fig: snr}.}
\label{fig:B_ni_Te_beta_3D}
\end{figure}

For notational clarity in the derivation and analysis that follows, we adopt a shorthand in which the explicit dependence on galactic coordinates ($R,Z$) is dropped. Additionally, we replace the ``total'' subscript with a ``0'' subscript to denote equilibrium values at a specific location. Thus, we have the mappings: $n_{e,\text{total}} \rightarrow n_{e0}$, $n_{i,\text{total}} \rightarrow n_{i0}$, $B_{0,\text{total}} \rightarrow B_0$, and $T_{e,\text{total}} \rightarrow T_{e0}$. For the positron fraction, we simply adopt the notation $p_{\text{total}} \rightarrow p$. This convention streamlines the fluid equations while preserving their spatial dependence implicitly.

\subsection{Quantitative Characterization of Galactic Profiles}
\label{sec:spatial_profiles}

To evaluate the location dependent KdV coefficients and the resulting KA soliton properties, we visualize how the background plasma parameters vary across our structured ISM model. Using the formulations in 
Sec. \ref{subsec:multi-component}--\ref{subsec:total_ISM_Ion_Density}, 
we construct two-dimensional (2D) Galactic maps (Fig.~\ref{fig:B_ni_Te_beta_3D}) and detailed one-dimensional (1D) midplane profiles ($Z=0$; Figs.\ref{fig: swb} and \ref{fig: snr}). 
These profiles capture the combined effects of the WIM and the embedded structures, serving as the environmental inputs for the soliton analysis presented in 
Sec.~\ref{gov_equations}.
While the HII region's impact is captured in the global 2D maps, we omit its corresponding 1D radial profiles for brevity. As these regions are modeled via simple Gaussian distributions for all plasma parameters, their 1D cross-sections lack the complex morphological features, such as shocked shells and interior plateaus, that characterize SWBs and SNRs. Consequently, our detailed 1D analysis focuses exclusively on the latter two structures to better illustrate the impact of sharp gradients and multi-component core-shell interactions on KA soliton dynamics.

In the global maps of Fig.~\ref{fig:B_ni_Te_beta_3D}, each embedded structure is outlined with a dashed magenta circle of radius $0.25\,$kpc. This adopted radius corresponds to the characteristic physical scale parameter used in our model: for the H\,II region, it represents the Gaussian width ($\sigma_{n,\mathrm{HII}} \approx 0.25\,$kpc), while for the SWB and SNR, it represents the shell radius ($R_{\mathrm{sh}} \approx 0.25\,$kpc). We clarify that these values are exaggerated compared to typical physical radii of these structures ($\sim10-50\,$pc) to ensure their gradients are visible against the multi-kiloparsec extent of the Galactic disk. Although the characteristic radius of $0.25\,$kpc marks the nominal scale of each structure, their spatial influence varies by type. For the H\,II region, the Gaussian density profile produces a diffuse enhancement that spreads significantly outside the $0.25\,$kpc circle. In contrast, the SWB and SNR are modeled with sharper shell and plateau functions (representing shock fronts), resulting in more localized structures whose steep gradients are confined relatively close to this nominal boundary.

\subsubsection{Galactic Background and H~II Regions}
Fig.~\ref{fig:B_ni_Te_beta_3D} presents the galactic 2D distribution of the plasma parameters in the $(R,Z)$ plane. All parameters are plotted on a $\log_{10}$ scale to capture the wide dynamic range of the ISM. The large scale background follows the disk-halo morphology, with magnetic field strength ($B_0$, Fig.~\ref{fig:B_ni_Te_beta_3D}a) and density ($n_{i0}$, Fig.~\ref{fig:B_ni_Te_beta_3D}b) decreasing gradually with Galactic height and radius. In the 2D maps, the background magnetic field is maintained around $\log_{10} (B_0/\mu\mathrm{G}) \approx 0.85$, which corresponds to a physical field strength of $B_0 \approx 7\,\mu\mathrm{G}$ , consistent with standard WIM values. Similarly, the background electron temperature (Fig.~\ref{fig:B_ni_Te_beta_3D}c) remains near $\log_{10} (T_{e0}/K) \approx 3.9$ ($T_{e0} \approx 8000$ K). The parameters adopted for the WIM background and the structural enhancements of the embedded astrophysical environments are given in Tables~\ref{tab:param_background} and \ref{tab:profiles_parameters} 
respectively.

Embedded within this smooth background at $(R,Z) \approx (4,0)$ kpc is the H\,II region. In the 2D maps, it appears as a distinct, localized perturbation: a diamagnetic cavity in the magnetic field (Fig.~\ref{fig:B_ni_Te_beta_3D}a) coincident with a density enhancement (Fig.~\ref{fig:B_ni_Te_beta_3D}b) and a modest thermal bump (Fig.~\ref{fig:B_ni_Te_beta_3D}c). The high thermal pressure and reduced magnetic field within the H\,II region drive the local plasma beta significantly above unity ($\beta > 1$). In Fig.~\ref{fig:B_ni_Te_beta_3D}d, this high-$\beta$ regime corresponds to the region masked in cyan. The implications of this $\beta$ distribution for soliton existence are analyzed in  
Sec.~\ref{sec:discussion}.

\subsubsection{Stellar-Wind Bubbles}
\begin{figure}[!htbp]
\centering
\includegraphics[width=0.45\textwidth,height=0.42\textwidth]{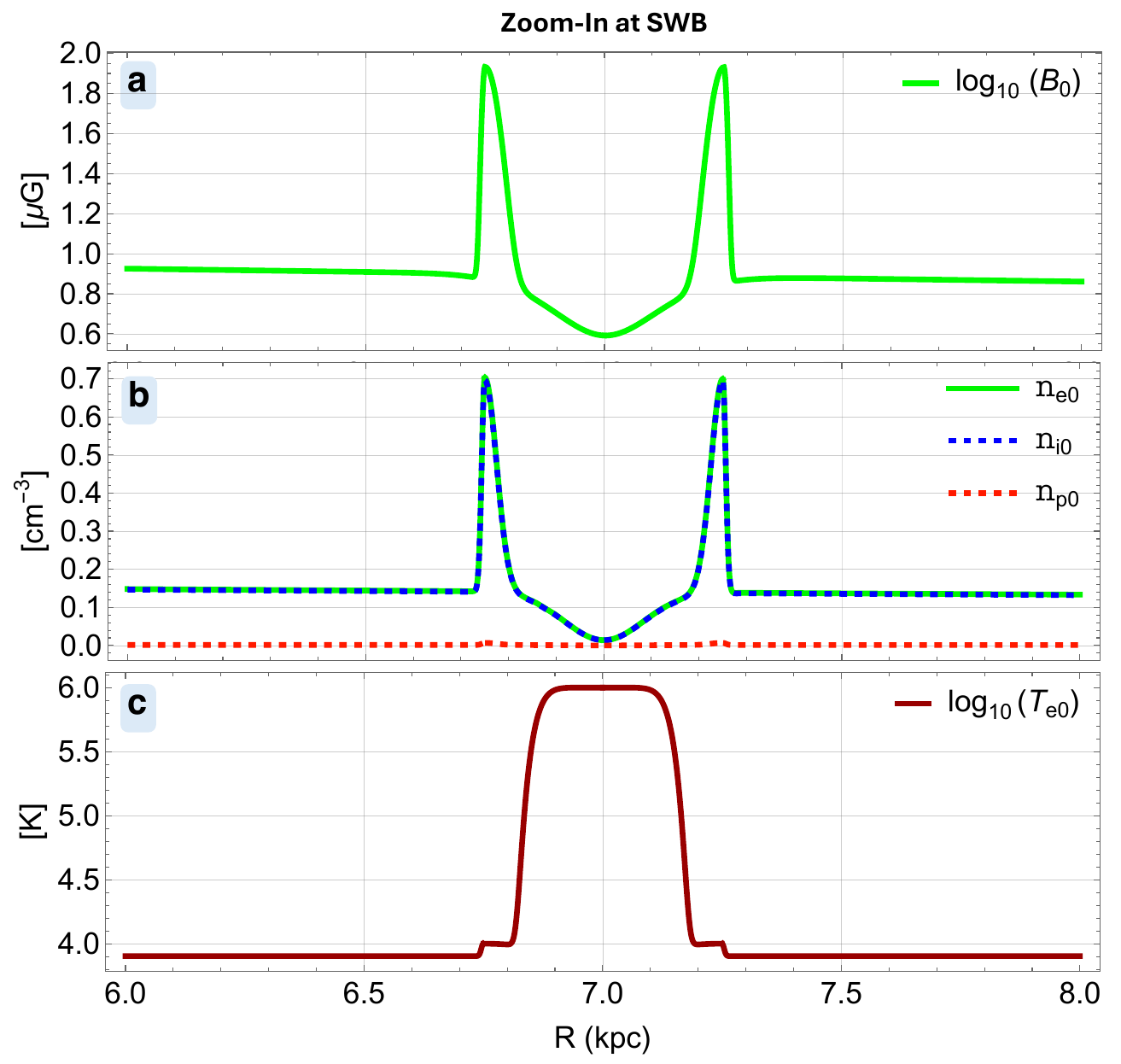}
\caption{Detailed 1D profiles of ISM parameters through the center of the SWB.
(a) Total magnetic field $B_0$ ($\mu$G), showing slight compression at the shell but minimal interior perturbation.
(b) Number densities ($\text{cm}^{-3}$) for electrons ($n_{e0}$, solid green), ions ($n_{i0}$, dotted blue), and positrons ($n_{p0}$, dashed red). Note the deep central cavity where ion density drops significantly.
(c) Electron temperature $T_{e0}$ (K), exhibiting the characteristic ``flat-top'' plateau of the hot shocked wind ($T \sim 10^6$ K).}
\label{fig: swb}
\end{figure}

The SWB, centered at $(R,Z) \approx (7,0)$ kpc, introduces more complex, fine-scale variations that are difficult to discern in the global 2D maps. To resolve these features, we present spatially coincident 1D profiles of the SWB parameters in Fig.~\ref{fig: swb}.
The magnetic structure of the SWB exhibits a distinct shell morphology in the global 2D map (Fig.~\ref{fig:B_ni_Te_beta_3D}a). This is quantitatively resolved in the 1D profile (Fig.~\ref{fig: swb}a) as characteristic twin peaks at the shell boundaries, resulting from the compression of frozen-in field lines by the expanding wind. These peaks enclose a central region where the field is stretched and slightly diluted.

The density profile (Fig.~\ref{fig: swb}b) follows a similar but more pronounced ``hole-and-shell'' morphology. While visible as a small, low-density cavity in the global maps (Fig.~\ref{fig:B_ni_Te_beta_3D}b), its internal structure is best resolved in the 1D profile, which shows a deep central cavity where the electron/ion ($n_{e0}\approx n_{i0}$ due to charge neutrality) density drops to near-vacuum levels ($\sim 0.01$ cm$^{-3}$), surrounded by a sharp, compressive shell. Note that unlike the log-scale global maps, these 1D density profiles are plotted on a linear scale to clearly illustrate the depth of the evacuation.

In contrast, the temperature profile (Fig.~\ref{fig: swb}c) exhibits the inverse behavior, characterized by the ``flat-top'' plateau of the hot shocked wind ($T \sim 10^6$ K) derived from the super-Gaussian model. This intense internal heating drives the local plasma beta above unity. In the global $\beta$ map (Fig.~\ref{fig:B_ni_Te_beta_3D}d), this feature appears as a central cyan mask, while the detailed 1D profile (see Fig. \ref{fig:SWB_1d_beta_amp_width}a) highlights the specific radial extent where $\beta > 1$.

\begin{figure}[!htbp]
\centering
\includegraphics[width=0.45\textwidth,height=0.4005\textwidth]{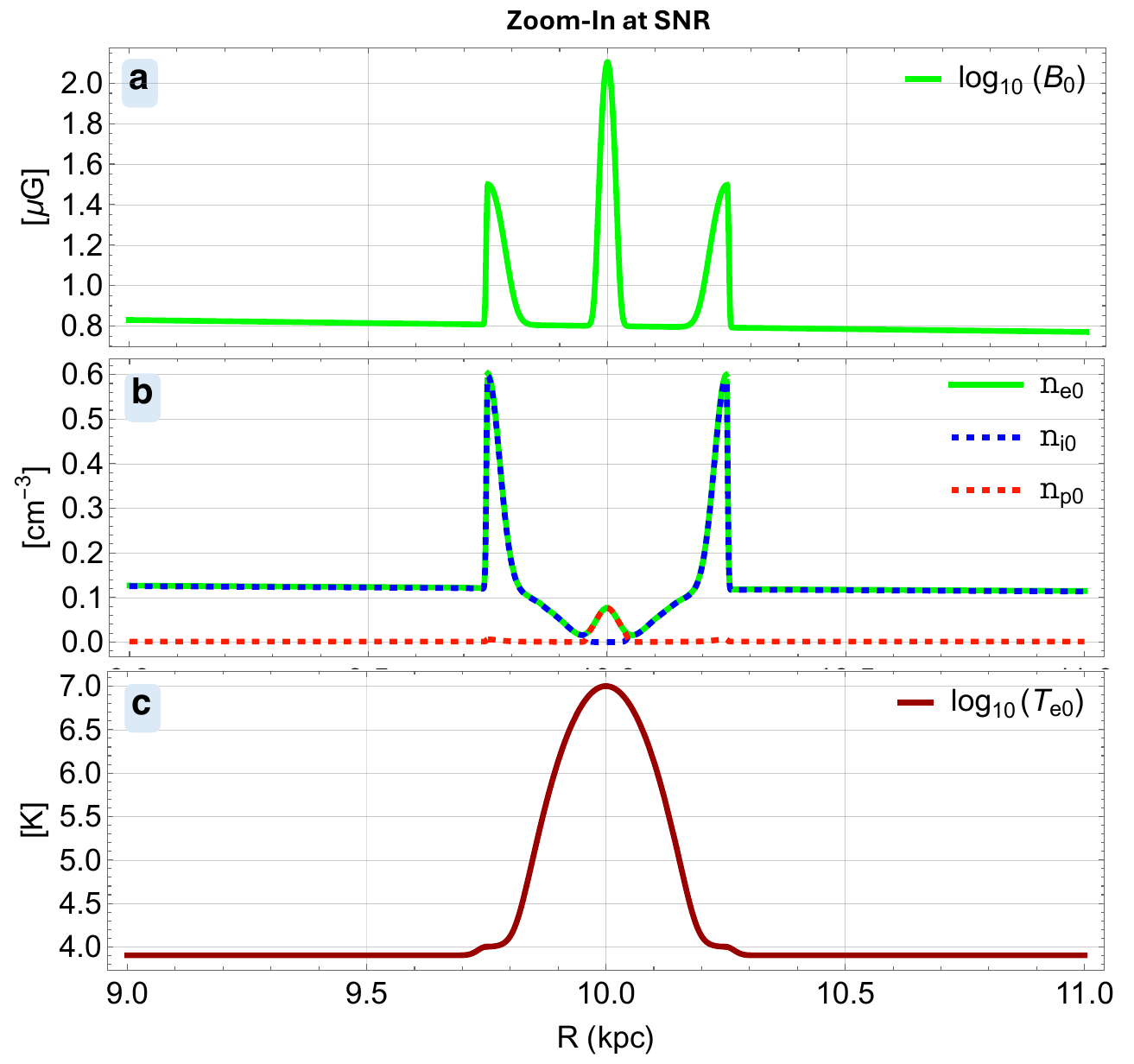}
\caption{Detailed one-dimensional profiles of ISM parameters through the center of the SNR.
(a) Total magnetic field $B_0$ ($\mu$G), highlighting the twin peaks of the shock-compressed shell and the central spike corresponding to the PWN.
(b) Multi-species densities ($\text{cm}^{-3}$): The ion density ($n_{i0}$, dotted blue) forms a shell but vanishes in the core, while the positron density ($n_{p0}$, dashed red) rises centrally, creating a pair-dominated interior.
(c) Electron temperature $T_{e0}$ (K), showing intense shock heating at the shell.}
\label{fig: snr}
\end{figure}
\subsubsection{Supernova Remnants}
The SNR, located at $(R,Z) \approx (10,0)$ kpc, represents the most extreme magnetic and thermal environment in our model. Its intricate multi-component structure creates sharp gradients that appear as bright rings (density/temperature) and central spikes (magnetic field) in the 2D maps of Fig.~\ref{fig:B_ni_Te_beta_3D}. These features are quantitatively analyzed via the 1D profiles in Fig.~\ref{fig: snr}.
The 1D magnetic profile (Fig.~\ref{fig: snr}a) is characterized by a shell-like enhancement at the shock front ($\sim 30\,\mu$G, $\log_{10} (B_0/\mu\mathrm{G}) \approx 1.5$) enclosing a cavity. Crucially, the profile also features a pronounced central spike, reaching values $\ge 120\,\mu$G, which corresponds to the strong magnetic field associated with the PWN core. The intrinsic neutron star magnetic field is not explicitly shown, as it decays rapidly with radius and becomes negligible on the kiloparsec scales resolved here.

The density morphology (Fig.~\ref{fig: snr}b) is complex and species dependent. Plotted on a linear scale, the total electron/ion density exhibits twin peaks corresponding to the swept-up shell, followed by a decline toward the interior. For the majority of the remnant, the electron (solid green) and ion (dotted blue) profiles are merged, confirming the maintenance of charge neutrality ($n_{e0} \approx n_{i0}$). However, in the deep core, the electron density rises again due to the injection of relativistic pairs. A key feature is the behavior of the ion density (blue dotted line), which drops to negligible values near the center ($n_{i0} \approx 0$), confirming the formation of an ionic cavity. Conversely, the positron density (red dotted line) peaks in this core region, separating from the ion profile and reflecting the pair-dominated nature of the PWN.

The SNR dominates the thermal map (Fig.~\ref{fig:B_ni_Te_beta_3D}c) with a massive temperature spike. The 1D profile (Fig.~\ref{fig: snr}c) shows this temperature rising sharply at the shock front to $>10^6$ K and remaining elevated throughout the interior, consistent with the Sedov-Taylor blast wave physics described in Sec.~\ref{subsubsec:SNR}. 
Unlike the denser, collisional H\,II region where viscosity may damp the cascade, the shock heated plasma in the SNR shell is effectively collisionless, permitting the turbulent cascade to extend down to the kinetic scales required for KAW generation.
Throughout this model, we assume the positrons are in thermal equilibrium with the electrons ($T_{p0} \approx T_{e0}$). 
Although the central PWNe is known to inject additional relativistic particle energy, this contribution is conservatively neglected in our model. PWNe thermal heating is typically confined to sub-parsec scales and is secondary to the broader, more intense shock heating at the shell, which dominates on the kpc-resolved grid \citep{TEMIMLATETIME2015}.
The distribution of plasma $\beta$ (Fig.~\ref{fig:B_ni_Te_beta_3D}d) at SNR exhibits a highly layered morphology. The shock heated shell, where $\beta > 1$,  appears as a cyan ring, while the deep interior core, where magnetic field is strong and $\beta < m_e/m_i$, is marked by a green dot. The detailed correspondence between these $\beta$ features and the allowable soliton regimes is discussed in Sec.~\ref{sec:discussion}.

\section{Governing Fluid Equations for KA Solitons \label{gov_equations}}
Having established the large scale physical environment of the ISM, we now formulate the fluid equations that govern the dynamics of KA solitons. 
Our model assumes a collisionless, magnetized electron-positron-ion (e-p-i) plasma, consistent with conditions in the WIM. 
The dynamics are described within a local Cartesian frame $(x,z)$, where the equilibrium plasma parameters are determined by our multi-component ISM model at a specific galactic location $(R, Z)$. 
In this frame, positive ions are treated as a fluid, while the inertialess electrons and positrons follow superthermal kappa ($\kappa$) distributions, a common feature in space and astrophysical plasmas shaped by energetic processes.

To generalize the fluid equations, all physical quantities are normalized by characteristic scales that are themselves functions of the local ISM parameters. This ensures our results explicitly depend on the galactic location. Time is normalized by the ion-acoustic transit time across a gyroradius:  $t=t'\,V_A/\rho_i$; velocities by the Alfvén speed:  $v_{ix,iz}=v_{ix,iz}'/V_A$; potentials by characteristic energy scale $(k_BT_{e0}/e)$: $(\phi,\psi)=(\phi',\psi')/(k_BT_{e0}/e)$; spatial coordinates by the ion gyroradius: $x,z=(x',z')/\rho_i$; and densities by their local equilibrium values: $n_j=n_j'/n_{j0}$. The key characteristic parameters are thus defined locally as: $V_A= {B_0}/{\sqrt{4\pi n_{i0} m_i}}$, $\Omega_i= {e B_{0}}/{m_i c}$, $C_s= \sqrt{{k_B T_{e0}}/{m_i}}$, $\rho_i= {C_s}/{\Omega_i}$, $\beta = {8\pi n_{i0} k_B T_{e0}}/{B_{0}^2}$. 
In this framework, the dynamics of KA solitons in a low-$\beta$ regime are governed by the dimensionless equations presented below (Eqs.~\ref{eq:continuity_norm}--\ref{eq:nl_kappa_norm}), where the dimensionless parameters explicitly depend on galactic location via our multi-component ISM model. This provides a direct path to studying KA solitons within specific, observable astronomical structures.

We consider ions (proton) as a cold fluid, neglecting their thermal pressure. This approximation is justified for the WIM, where the plasma beta is low and the ion thermal velocity is significantly smaller than the Alfvén speed. Since our focus is on the nonlinear evolution of KA solitons, whose dynamics are primarily governed by magnetic and inertial effects, the cold ion assumption simplifies the model without compromising physical accuracy \citep{SINGHSelfConsistent2025}: 
\begin{equation}
\frac{\partial n_{i}}{\partial t}+\frac{\partial (n_{i}v_{ix})}{\partial x}+\frac{\partial (n_{i}v_{iz})}{\partial z}=0,
\label{eq:continuity_norm}
\end{equation}
\begin{equation}
v_{ix}=-\Lambda \frac{\partial^{2} \phi}{\partial x \partial t},
\label{eq:momentum_x_norm}
\end{equation}
\begin{equation}
\frac{\partial v_{iz}}{\partial t}+v_{ix}\frac{\partial v_{iz}}{\partial x}+v_{iz}\frac{\partial v_{iz}}{\partial z}=-\Lambda \frac{\partial \psi}{\partial z},
\label{eq:momentum_z_norm}
\end{equation}
\begin{equation}
\Lambda\frac{\partial^{4}(\phi-\psi)}{\partial x^{2}\partial z^{2}}= \frac{\partial^{2} n_{i}}{\partial t^{2}}+\frac{\partial^{2}(n_{i}v_{iz})}{\partial z \partial t},
\label{eq:coupling_norm}
\end{equation}
where $\Lambda=\beta/2$. 
The equilibrium charge neutrality condition, expressed in terms of the density ratios as  $\delta_{ei}=1+p,$
is determined explicitly by the local positron fraction $p$ (as defined in Eq.~\ref{p-total}), where $\delta_{ei} \equiv n_{e0}/n_{i0}$. 
The normalized densities of the superthermal electrons and positrons, 
in response to the parallel potential $\psi$ of KA solitons, are given by \citep{SinghKourakis2019}:
\begin{equation}\label{eq:nl_kappa_norm}
n_{l}=\left[1 - {R_l}\,\frac{\psi}{\kappa_{l}-\tfrac{3}{2}}\right]^{-\kappa_{l}+\tfrac{1}{2}}
\;\;\approx\;\; 1 + c_{1l}\psi + c_{2l}\psi^2,
\end{equation}
where $l \in \{e,p\}$ denotes electrons ($e$) and positrons ($p$), with $R_{e}=+1,\quad R_{p}=-\alpha$, and the expansion coefficients are expressed as
\begin{align}
c_{1l} &= R_{l}\,\frac{\kappa_{l}-{1}/{2}}{\kappa_{l}-{3}/{2}}, &
c_{2l} &= R_{l}^{2}\,\frac{\kappa_{l}^{2}-{1}/{4}}{2(\kappa_{l}-{3}/{2})^{2}}, \nonumber
\end{align}
where $\kappa_{l}$ denotes the spectral index quantifying the superthermality of species $l$, 
and $\alpha \equiv T_{e0}/T_{p0}$ represents the electron-to-positron temperature ratio.
This set of self-consistent equations forms the basis for the derivation of the KdV equation in the subsequent sections.



\section{Derivation of the KdV equation and its solution \label{kdv-derivation}}

The KdV equation is derived using the reductive perturbation method, and defining the independent stretched coordinate system $\xi$ and $\tau$ as:
\begin{equation}
    \xi=\epsilon^\frac{1}{2}(l_{x}x+l_{z}z-\lambda t), \quad \tau=\epsilon^{\frac{3}{2}}t,
    \label{stretched-coord}
\end{equation}
where $\lambda$ is the phase velocity of the KAWs, $\epsilon$ is a small expansion parameter ($0<\epsilon\ll 1$) characterizing the weak nonlinearity. The geometric parameters $l_{x}$ and $l_{z}$ represent the direction cosines in the $x-z$ plane, satisfying condition $l_{x}^{2}+l_{z}^2=1$. 
The dependent variables can be expanded as a power series in $\epsilon$:
\begin{equation}
    S=\sum_{q=1}^\infty \epsilon^qS^{(q)}, \quad \rm{and} \quad \phi=\sum_{q=1}^\infty \epsilon^{q-1}\phi^{(q)},
    \label{perturbations}
\end{equation}
\noindent
where $S=(v_{ix},v_{iz},\psi)$. Note that while Eq.~(\ref{ion_density_derivation}) defines the dimensional equilibrium ion density $n_{i0}$ used to construct the global maps, the variable $n_i$ in the expansions below refers to the dimensionless ion density normalized to this local background. Consequently, the equilibrium state corresponds to $n_i = 1$ and $\psi = 0$. According to plasma approximation, the normalized expression can be written as: $n_i=(1+p)\,n_e-p\,n_p.$
Using Eq. (\ref{eq:nl_kappa_norm}) in above equation, we obtain 
\begin{equation}\label{ni=psi}
n_i=1+a_{1}\psi+a_{2}\psi^2+...
\end{equation}
where $a_1=(1+p)\,c_{1e}+p\,c_{1p}$ and $a_2=(1+p)\,\delta_{ei}c_{2e}-p\,c_{2p}$. 
Using Eqs. (\ref{stretched-coord}), (\ref{perturbations}) and (\ref{ni=psi}) in  Eqs. (\ref{eq:continuity_norm})-(\ref{eq:coupling_norm}) and comparing the coefficients of lowest powers of $\epsilon$, we obtain the first order governing Eqs. 
(\ref{eq16})-(\ref{eq19}), given in Appendix \ref{App.-A}.  
On solving these first-order governing equations, we obtain the biquadratic dispersion equation
\begin{equation}\label{eq20}
a_{1}\lambda^{4}-(a_{1}+\beta)\,l_{z}^{2}\lambda^{2}+\beta l_{z}^{4}=0. \nonumber
\end{equation}
On simplification of the above biquadratic equation, we obtain two different roots corresponding to KA mode and ion acoustic mode which are given respectively as: $\lambda^2=l_{z}^{2}$ and $\lambda^2={\beta l_{z}^{2}}/{a_1}$.
The next order of $\epsilon$ yields the second order evolution Eqs. (\ref{eq23})-(\ref{eq26}),
given in Appendix \ref{App.-B}. 
Eliminating the second order perturbed quantities from these equations, the following KdV equation is obtained:
\begin{equation}\label{eq27}
\frac {\partial\psi^{(1)}}{\partial \tau}+ P \psi^{(1)} \frac{\partial\psi^{(1)}}{\partial \xi}+ Q \frac{\partial^{3}\psi^{(1)}}{\partial \xi^{3}}= 0,
\end{equation}
where $P=-a_{1}l_{z}$ is the nonlinear coefficient and $Q=-\frac{l_{x}^{2}l_{z}\beta}{2(a_{1}-\beta)}$
is dispersion coefficient. 
The stationary solution of the KdV equation (\ref{eq27}) is obtained by introducing the traveling wave transformation 
$\eta = \xi - U\tau$. The resulting soliton solution is given by \citep{sainiDustKineticAlfven2015,SAINIDust2017,Geetika22}:
\begin{equation}\label{eq30}
\psi^{(1)} = \psi_{0}\,\mathrm{sech}^{2}\!\left(\frac{\eta}{W}\right),
\end{equation}
where $U$ denotes the soliton speed, $\psi_{0} = \tfrac{3U}{P}$ represents the maximum amplitude, and 
$W = \sqrt{\tfrac{4Q}{U}}$ characterizes the soliton width.
The analytical solution given in Eq.~(\ref{eq30}) describes the local structure of a KA soliton in a homogeneous plasma. However, in our realistic ISM model, the nonlinear coefficient $P$ and dispersive coefficient $Q$, and consequently the soliton amplitude $\psi_0$ and width $W$, are not constants. Instead, they become spatially dependent functions determined by the local values of $\beta(R,Z)$, density $n_{i0}(R,Z)$, and positron fraction $p(R,Z)$ at each galactic coordinate. By evaluating these derived quantities across our global grid, we can now map the regions where soliton formation is physically permissible and characterize their properties within the heterogeneous ISM environment.

\section{Discussion}
\label{sec:discussion}

A central result of our analysis is the identification of exclusion zones (EZs) where KA solitons cannot exist.
These EZs correspond to areas where the local plasma $\beta$  violates the stability condition $m_e/m_i \ll \beta \ll 1$. In Fig.~\ref{fig:SWB_1d_beta_amp_width}a and \ref{fig:SNR_1d_beta_amp_width}a, the green lines represent $\log_{10}\beta$, with dotted blue and red lines marking $\beta=1$ ($\log_{10}\beta=0$) and $\beta=m_e/m_i$ ($\log_{10}\beta\simeq -3.3$), respectively. Specifically, the EZs arise in regimes where $\beta > 1$ (masked in cyan in Fig.~\ref{fig:B_ni_Te_beta_3D}d and shaded red in the 1D profiles in Figs.~\ref{fig:SWB_1d_beta_amp_width}a and \ref{fig:SNR_1d_beta_amp_width}a) or $\beta \ll m_e/m_i$ (masked in green in Fig.~\ref{fig:B_ni_Te_beta_3D}d and shaded yellow in the 1D profile of Fig.~\ref{fig:SNR_1d_beta_amp_width}a).  
The EZs are predominantly located at the dense H\,II regions (where the high-$\beta$ regime dominates the interior as well as extends significantly beyond the nominal structure radius), the hot interiors of both SWBs and SNRs ($\beta > 1$), and the deep, magnetically dominated cores of SNRs ($\beta \ll m_e/m_i$). The compressed shells of both SWBs and SNRs largely maintain a favorable $\beta$ ($m_e/m_i < \beta < 1$), positioning these shock-bounded structures as prime candidates for the formation of KA solitons.

\begin{figure}[!htbp]
\centering
\includegraphics[width=0.45\textwidth,height=0.34005\textwidth]{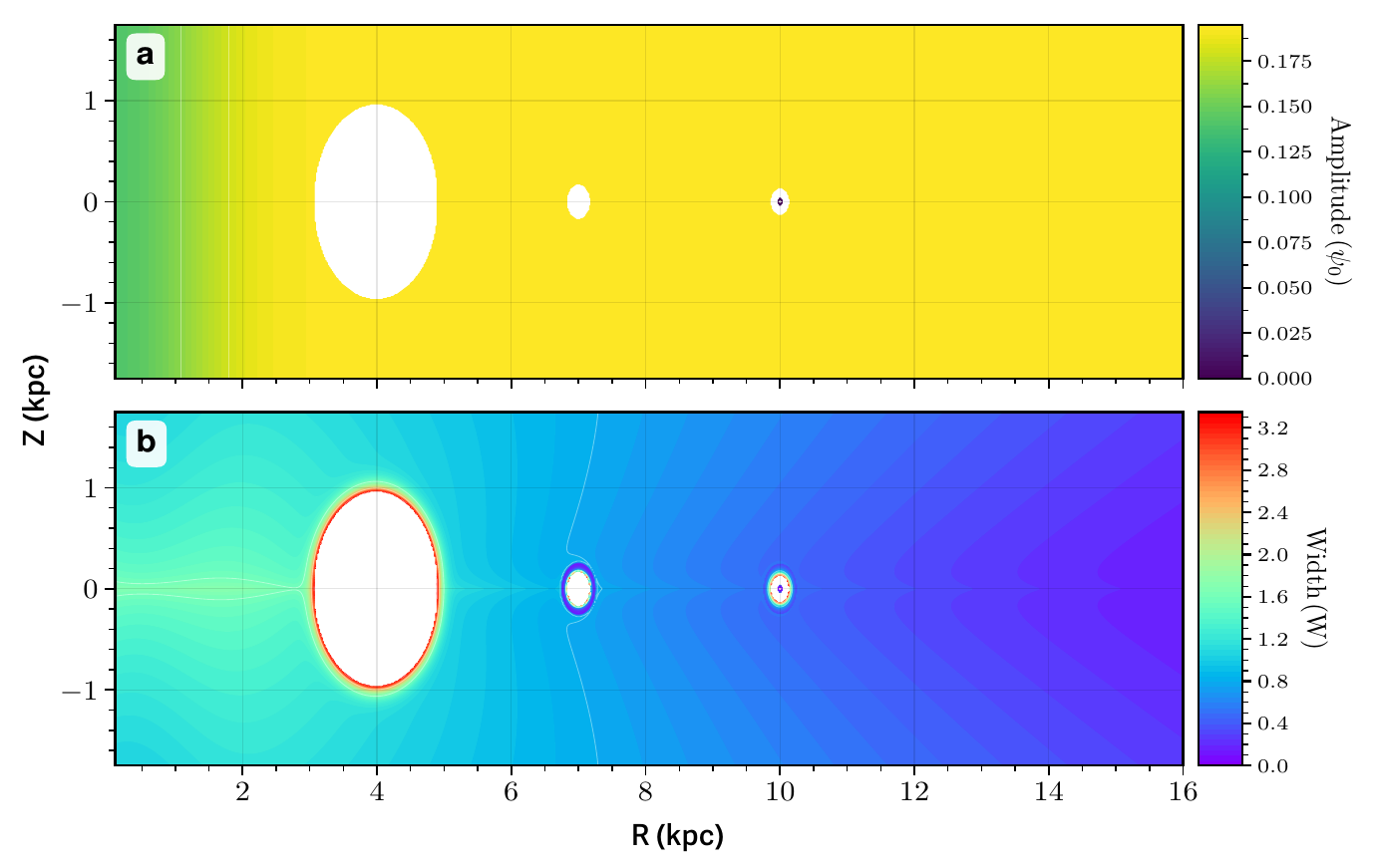}
\caption{Two-dimensional Galactic maps of KA soliton (a) amplitude $\psi_0$ and (b) width W, at $\kappa_e=150, \kappa_p=100$, $\theta=85^\circ$, $U=-0.06$ and $\sigma=1$. The parameters for the background (WIM) and the analytical profiles of embedded structures are given in Tables~\ref{tab:param_background} and ~\ref{tab:profiles_parameters} respectively.
}
\label{fig: full_map_amp_width}
\end{figure}

While the EZs mapped in our analysis are formally defined by the $\beta$ limits ($m_e/m_i \ll \beta \ll 1$), this limit may be intrinsically linked to the physical regimes where the turbulent cascade is disrupted or fundamentally altered. Thus, the formation of KA solitons requires the satisfaction of two coupled conditions: (1) a compatible plasma $\beta$ window where the soliton solution is mathematically valid, and (2) a turbulent cascade capable of penetrating to ion-kinetic scales ($k_\perp \rho_i \sim 1$) to seed the finite-amplitude fluctuations required for nonlinear steepening. 

In $\beta > 1$ environments, thermal pressure dominates, promoting compressive fluctuations that are subject to rapid collisionless damping. This regime inhibits the anisotropic Alfv\'enic cascade necessary for KAWs, as compressive modes damp more rapidly.
In partially ionized H\,II regions, additional damping from ion-neutral friction further cuts off the cascade before it reaches kinetic scales, preventing the generation of seed perturbations for solitons. Consequently, these high-$\beta$ EZs represent regions where the energy flux is likely diverted into ion heating or other dissipative channels rather than forming coherent KAW solitons, marking boundaries between different kinetic dissipation pathways in the ISM.

Notably, narrow transition annuli exist within SNRs, where $m_e/m_i < \beta < 1$ and the cascade can robustly reach kinetic scales. These favorable zones are spatially sandwiched between the high-$\beta$ interior (masked in cyan) and the low-$\beta$ core (masked in green), as shown in the inset of Fig.~\ref{fig:B_ni_Te_beta_3D}d. In the 1D profiles (Fig.~\ref{fig:SNR_1d_beta_amp_width}a), this transition manifests as the distinct green-shaded region separating the central yellow core from the surrounding red regions. 
The plasma in this annulus is effectively collisionless, where viscosity cannot terminate the cascade, compelling the turbulence to extend down to the ion gyroradius scale. Thus, this annulus represent a unique site where both the spectral requirement (a robust cascade to kinetic scales) and the parametric requirement (a compatible $\beta$ window) can simultaneously be met, enabling soliton formation amid intense turbulence. 
Physically, this zone marks the interface where the star-injected magnetic flux and turbulence equilibrate with the surrounding thermal environment. Unlike the outer blast wave, this region is governed by the central compact object, implying that the KA solitons predicted here are driven by the balanced injection of magnetic energy and plasma from the pulsar wind rather than the extreme compression of the forward shock.
The EZs therefore do not imply the absence of turbulence or wave activity; instead, they indicate a regime transition where turbulence persists but does not organize into coherent, anisotropic KA solitons, and the combination of large $\beta$ and strong compressibility is hostile to their formation.

\subsection{Amplitude and Width}

Fig.~\ref{fig: full_map_amp_width} provides a comprehensive galactic map of the KA soliton amplitude ($\psi_0$) and width ($W$). The amplitude (Fig.~\ref{fig: full_map_amp_width}a) remains remarkably constant across much of the Galaxy, hovering around $\psi_0 \sim 2 \times10^{-1}$, with minimal variation near the embedded structures like H\,II regions, SWBs, and SNRs. This near constancy suggests that $\psi_0$ is largely insensitive to local variations in plasma $\beta$, magnetic field strength ($B_0$), or density gradients in our model, which assumes a fixed perturbation strength for soliton seeding. This behavior is therefore a diagnostic of the present leading-order KdV formulation: the amplitude is only weakly (or not explicitly) coupled to $\beta$ and $B_0$, so large-scale environmental changes cannot strongly imprint themselves on $\psi_0$. 
However, notable deviations occur near the Galactic center ($R \lesssim 2$ kpc), where $\psi_0$ decreases to $\sim 1.5 \times10^{-1}$, likely due to the denser, more turbulent environment suppressing large amplitude coherent structures.

\begin{figure}[!htbp]
\centering
\includegraphics[width=0.45\textwidth,height=0.4005\textwidth]{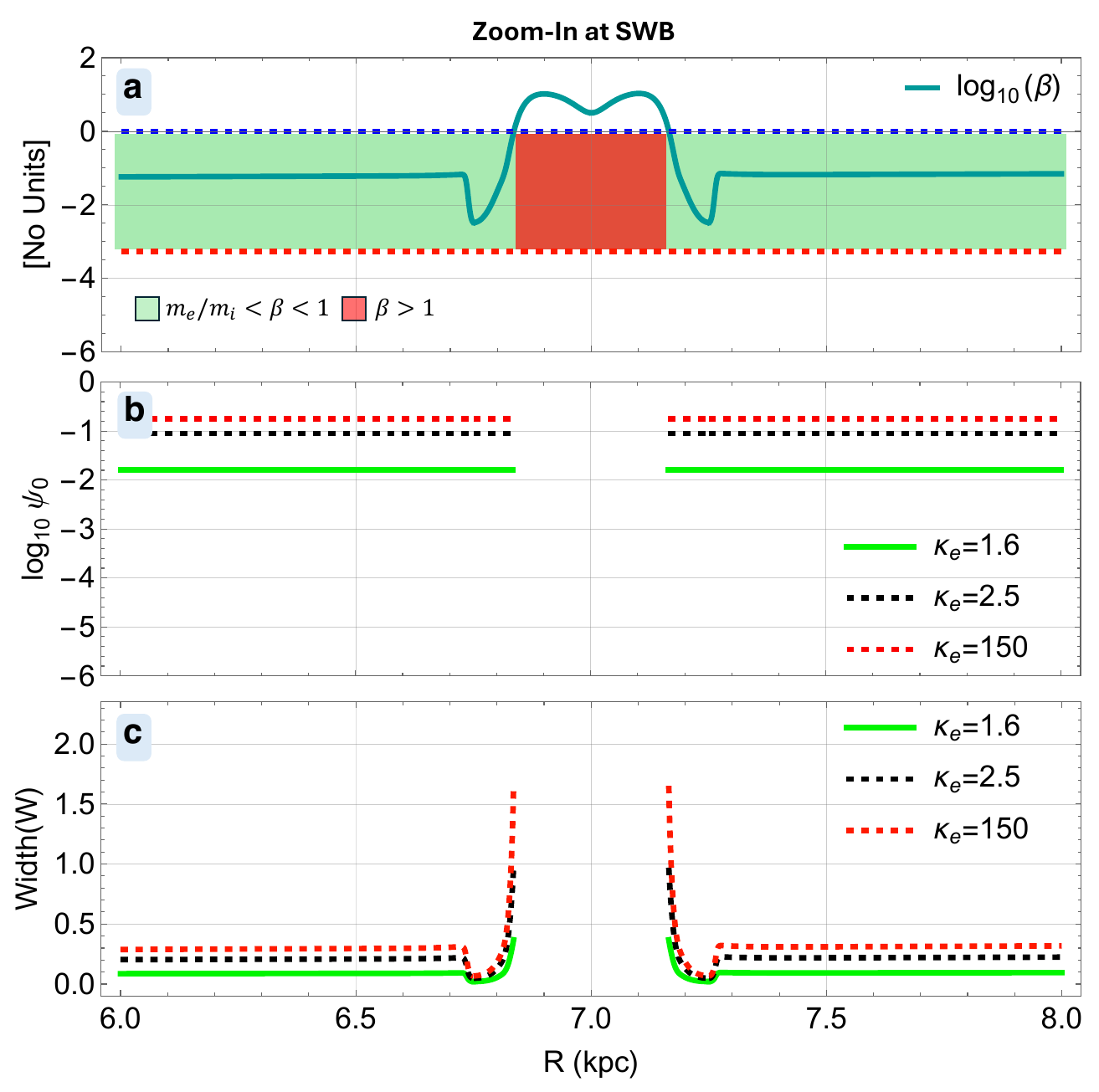}
\caption{One-dimensional profiles of plasma $\beta$ in (a), amplitude in (b) and width in (c) for $Z=0\,$kpc at SWB for same parameters as in Fig. \ref{fig: full_map_amp_width}.}
\label{fig:SWB_1d_beta_amp_width}
\end{figure}

The soliton width (Fig.~\ref{fig: full_map_amp_width}b) exhibits greater variability, with sharp increases (red patches) at the boundaries of EZs. These broad widths at EZ edges reflect enhanced dispersion from larger gradients in temperature and density. Within the KdV scaling ($W\propto \sqrt{Q/U}$), the approach to the EZ boundaries ($\beta\to 1$ or $\beta\to m_e/m_i$) can drive a rapid increase of the effective dispersive balance, producing strong broadening immediately before the soliton solution disappears. Narrower widths dominate the quiescent WIM, indicating stable, compact solitons in relatively uniform plasma regions. The EZs themselves appear as white voids, reinforcing the $\beta$-driven prohibition on soliton existence. Physically, this implies that solitons approaching a high-$\beta$ shell or an ultra low-$\beta$ cavity are expected to de-localize and disperse into a more linear wave train as the nonlinearity--dispersion balance fails.

\begin{figure}[!htbp]
\centering
\includegraphics[width=0.45\textwidth,height=0.4005\textwidth]{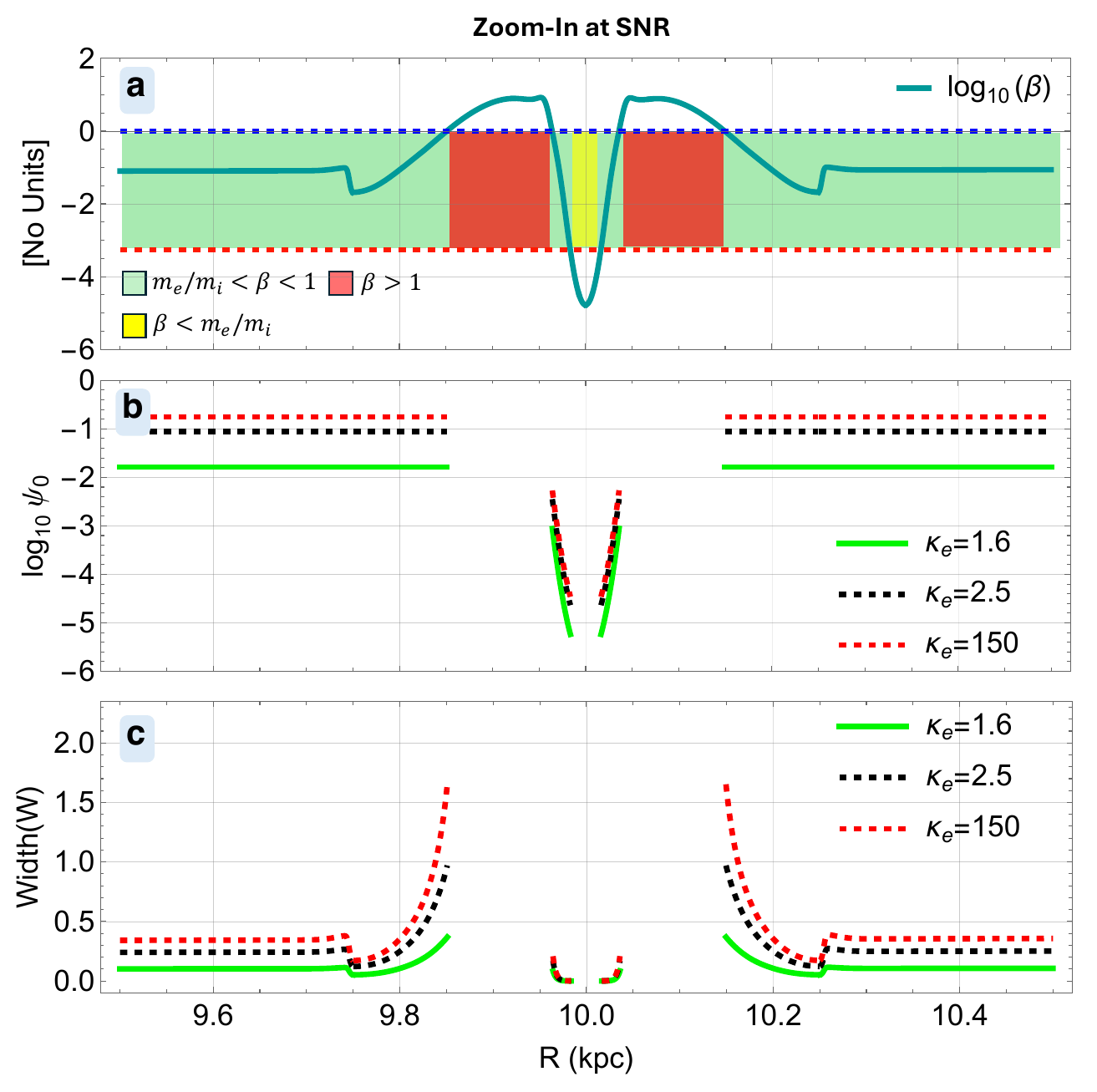}
\caption{One-dimensional profiles of plasma $\beta$ in (a), amplitude in (b) and width in (c) for $Z=0\,$kpc at SNR for same parameters as in Fig. \ref{fig: full_map_amp_width}.}
\label{fig:SNR_1d_beta_amp_width}
\end{figure}

Figs.~\ref{fig:SWB_1d_beta_amp_width}b,c and \ref{fig:SNR_1d_beta_amp_width}b,c show the 1D profiles of amplitude and width at SWB and SNR respectively. It is observed that for both SWB and SNR, the width shows a subtle dip just before the EZ boundary, followed by a sharp rise at the edge. This dip likely results from localized density and magnetic field enhancement at the shells of SWB and SNR, decreasing dispersion effects temporarily. The dip in width at SWB and SNR also correlates with dip in plasma beta at same $R-$coordinate (Figs.~\ref{fig:SWB_1d_beta_amp_width}a and \ref{fig:SNR_1d_beta_amp_width}a). The spikes in $W$ near the edges of SWB and SNR corresponds to the larger width regions manifested by red patches in Fig. \ref{fig: full_map_amp_width}.

A narrow annulus of reduced amplitude and width exist within the SNR interior. This feature corresponds precisely to the transition annulus identified in the $\beta$ stability maps, where $m_e/m_i < \beta < 1$. While the previous analysis identified this zone as the favorable zone for KA soliton excitation, the amplitude map reveals that these structures are physically constrained by unique thermodynamic environment. In the 1D amplitude profiles (Fig.~\ref{fig:SNR_1d_beta_amp_width}b), the narrow annulus manifests as localized patches of reduced amplitude on either side of $R=10\,$kpc. The discontinuity or the break in the curves corresponds to the ultra low-$\beta$ EZ (the PWN core), where the KA solitons does not exist. Flanking this exclusion zone, the visible line segments (patches) correspond to the favorable transition annulus ($m_e/m_i < \beta < 1$). Interestingly, while solitons are permitted here, they exhibit significantly lower amplitudes ($\psi_0 \ll 2 \times10^{-1}$) compared to the ambient ISM. Since the intense PWN magnetic field has largely decayed to background levels at the outer edge of this annulus, this suppression is unlikely to be magnetic in origin. Instead, it arises from the unique thermodynamic state of this annulus. The sharp local gradients in ion density and electron temperature, characteristic of the cavity-shell interface alter the balance between the nonlinear and dispersive coefficients. This creates a specific parameter regime where the resulting soliton potentials are valid but naturally compact and small-amplitude, distinct from the robust structures supported by the more homogeneous ambient ISM.

The influence of electron superthermality on soliton morphology is shown in Figs.~\ref{fig:SWB_1d_beta_amp_width}b,c and \ref{fig:SNR_1d_beta_amp_width}b,c. Solid green lines represent highly superthermal electrons ($\kappa_e=1.6$), dotted black lines represent moderate superthermality ($\kappa_e=2.5$), and dotted red lines represent near-Maxwellian distributions ($\kappa_e=150$). Across all environments, we observe a consistent trend where both the soliton amplitude and width increase as the plasma becomes more Maxwellian. This behavior is explained by the ability of the local electric potential to organize the plasma particles. In the superthermal regime where $\kappa_e$ is low, the plasma contains a large population of very high speed electrons in the distribution tail. Because these particles are so energetic, it is much harder for a local electric potential to trap them or hold them back to create a large charge separation. Since these particles can easily escape the potential well, the plasma cannot support a large potential hill. Therefore, the maximum amplitude ($\psi_0$) is naturally limited at a lower value compared to a Maxwellian plasma. In contrast, thermal particles are slower and easier to organize into a coherent structure. This allows the plasma to sustain a much higher potential drop before the structure becomes unstable, resulting in the higher amplitudes seen in the near-Maxwellian cases.
This change in particle dynamics also regulates the soliton width. The excess energetic particles at low $\kappa_e$ enhance the nonlinear steepening of the wave. This strong nonlinearity acts to compress the soliton into a narrower and more compact structure. As the plasma thermalizes and $\kappa_e$ increases, this nonlinear compression effect weakens relative to the dispersive spreading. This allows the soliton to expand spatially, leading to the broader profiles observed in thermalized regions. This broadening becomes most prominent near the boundaries of the exclusion zones, where steep local gradients in temperature and density further increase the dispersive spreading, causing the soliton to spread before it reaches the stability threshold.

\section{SUMMARY AND CONCLUSIONS} 
This work has established a spatially dependent theoretical framework for the existence and propagation of nonlinear KA solitons within the multi-component ISM. By employing the reductive perturbation method to a magnetized electron-positron-ion plasma, we derived location dependent KdV equation that capture how macroscopic Galactic structures modulate micro-scale plasma dynamics. Our analysis identified distinct exclusion zones where the stability condition $m_{e}/m_{i} \ll \beta \ll 1$ is violated, effectively prohibiting soliton formation in high-$\beta$ H\,II regions and SWB cavities, as well as ultra low-$\beta$ cores near central pulsar wind nebulae (PWN). While the soliton amplitude remains relatively invariant across the Galactic disk, the spatial width exhibits significant broadening near these EZ boundaries due to enhanced dispersion from local environmental gradients, particularly in the complex morphologies of SWBs and SNRs.

To transition this framework from a theoretical model to a predictive observational tool, specific and quantifiable metrics must be evaluated. The model can be used to calculate expected scattering measures and scintillation bandwidth contributions from soliton populations along particular Galactic sightlines, providing a direct comparison with pulsar scintillation data. Similarly, our predictions for small-scale ($< 0.01$ pc) density power spectrum enhancements offer a feasible target for future high-resolution radio scattering observations or Very Long Baseline Interferometry (VLBI) studies. A primary challenge in this endeavor will be disentangling soliton signatures from the broader turbulent spectrum, which necessitates a focus on non-Gaussian statistics and polarization signatures unique to coherent, Alfvénic structures. At present, these connections between soliton width and the underlying electron $\kappa$-distribution should be framed as a testable hypothesis requiring forward-modeled observables rather than an immediate diagnostic for ISM thermodynamics. In this workflow, our Galactic maps serve as foundational admissibility layers that predict soliton EZs within bright, high-$\beta$ emission nebulae and soliton halos in the surrounding moderate-$\beta$ transition zones.

Beyond direct observation, this framework provides a physics based sub-grid model for galaxy-scale or cosmic-ray propagation simulations that cannot resolve ion-kinetic scales. By identifying exactly where coherent kinetic structures are likely to emerge and defining their characteristic scales, our model moves beyond the traditional assumption of homogeneous turbulent dissipation. In its present form, our model provides a map of allowed and forbidden zones for these waves. This admissibility layer is ready to be used in larger Galactic simulations to better understand how heat moves and how high-energy particles are scattered across the Galaxy. With the future addition of explicit seeding and damping physics, this can evolve into a quantitative sub-grid closure for environment-dependent transport. This demonstrates that the ISM is not merely a passive container for turbulence but an active modulator that carves out exclusion zones and tailors the properties of nonlinear phenomena.

Ultimately, the multi-component analytical ISM model developed here is highly versatile and can be extended to study a vast array of astrophysical plasma phenomena. The framework is well-suited for investigating various plasma instabilities, particle acceleration mechanisms, and wave-particle interactions at diverse sites, such as colliding wind binaries, magnetic reconnection regions, and high-energy shock interfaces. By establishing a direct link between macroscopic Galactic morphology and ion-kinetic scale dissipation, this study shifts the paradigm of the turbulent cascade's end-state from a universal process to a geographically complex one, with significant ramifications for our understanding of ISM thermodynamics and the interpretation of high-resolution astrophysical data.

\begin{acknowledgements}
M.S. gratefully acknowledges support from the Basic Scientific Research Fund for Central Universities, China (Grant No. 2682025CX094). S.L. acknowledges funding from the National Natural Science Foundation of China (Grant No. 12375103). N.S.S. acknowledges support from the Council of Scientific and Industrial Research (India) under the Emeritus Scientist scheme (Ref. No. 21/1195/25/EMR‑II).
\end{acknowledgements}

\bibliographystyle{aa}
\bibliography{References_Rev_One}

\appendix

\section{Model of H\,II Regions}
\label{subsubsec:HII_Region}

Superposed on the quiescent background are distinct regions of significantly altered density, which we model as localized perturbations using specific analytical forms. First, H\,II regions represent strong positive electron density enhancements, $n_{e,\text{H\,II}}$. These are islands of plasma, often visible as bright emission nebulae, generated by intense photoionizing radiation from central O- or B-type massive stars \citep{Haffner2009}. Since these regions form from the gravitational collapse of dense molecular clouds, their internal density is significantly higher than the ambient ISM \citep{PAGELComposition1979,DOPITATheoretical1986,BRINCHMANNNew2008,SHIRAZISTARS2014,LILinking2025a}. Using the 
analytical formalism for Localized structures introduced in the main text, 
Sec.~\ref{subsec:Analytical_formalism}, 
we model this contribution as a localized Gaussian enhancement:
\begin{equation}
n_{e,\mathrm{HII}}(R,Z) = \mathcal{G}(d_{\mathrm{HII}}; A_{n,\mathrm{HII}}, \sigma_{n,\mathrm{HII}}),
\end{equation}
where $d_{\mathrm{HII}} = \sqrt{(R - R_{1})^2 + (Z - Z_{1})^2}$ is the radial distance from the center $(R_1, Z_1)$ of the H\,II region. Here, $A_{n,\text{H\,II}}$ represents the peak amplitude of the density enhancement and $\sigma_{n,\,\rm{H\,II}}$ is the characteristic size.

In H\,II regions, the high thermal pressure of the hot, ionized gas drives an outward expansion. This expansion displaces the frozen-in field lines, creating a magnetic cavity with reduced field strength \citep{Spitzer1978,Ferriere2001}. We model this effect as a localized Gaussian depletion:
\begin{equation}
B_{0,\mathrm{HII}}(R,Z) = \mathcal{G}(d_{\mathrm{HII}}; A_{B,\mathrm{HII}}, \sigma_{B,\mathrm{HII}}),
\end{equation}
where $A_{B,\text{HII}}$ represents the amplitude of the magnetic depletion (negative), and $\sigma_{B,\,\rm{H\,II}}$ is the characteristic width of the magnetic cavity.

Inside H\,II regions, photoionization heating, where excess energy from ionizing photons is converted into kinetic energy of electrons, elevates the temperature above the background WIM \citep{DRAINE2011}.
Crucially, this temperature rise is limited by a thermostat effect, as temperature increases, cooling via collisionally excited forbidden lines rises exponentially, effectively limiting the equilibrium temperature typically near $10^4$\,K \citep{OSTERBROCK2006a}. We model this local thermal enhancement as:
\begin{equation}
T_{\mathrm{e,HII}}(R,Z) = \mathcal{G}(d_{\mathrm{HII}}; A_{T,\mathrm{HII}}, \sigma_{T,\mathrm{HII}}),
\end{equation}
where $A_{T,\text{HII}}$ is the temperature excess over $T_{e,\text{bkg}}$ and $\sigma_{T,\,\rm{H\,II}}$ defines the thermal width of the region. Note that, the H\,II regions are modeled with a total electron temperature of $10^4\,$K, representing a $25\%$ thermal enhancement over the $8,000\,$K WIM background.

\section{Model of Stellar-Wind Bubbles}
\label{subsubsec:SWB}

The powerful radiatively driven winds from massive stars excavate large, low-density cavities in the ISM, surrounded by dense shells of swept-up gas \citep{CASTORInterstellar1975,WEAVERInterstellar1977}. To model this morphology, we construct the electron density profile as the superposition of a central depletion and a shell enhancement:
\begin{multline}
n_{e,\mathrm{SWB}}(R,Z) = 
\mathcal{G}(d_{\mathrm{SWB}}; A_{n,\mathrm{cav,SWB}}, \sigma_{n,\mathrm{cav,SWB}})
\\+ 
\mathcal{S}(d_{\mathrm{SWB}}; A_{n,\mathrm{sh,SWB}}, R_{\mathrm{sh,SWB}}, \sigma_{n,\mathrm{in,SWB}}, \sigma_{n,\mathrm{out,SWB}}),
\end{multline}
where $d_{\mathrm{SWB}} = \sqrt{(R - R_{2})^2 + (Z - Z_{2})^2}$ is the radial distance from the bubble center ($R_2,Z_2$). The term $A_{n,\mathrm{cav,SWB}}$ is negative, representing the deep density minimum in the tenuous shocked wind.
For the shell component, $A_{n,\mathrm{sh,SWB}}$ and $R_{\mathrm{sh,SWB}}$ denote the peak density and shell radius, while $\sigma_{n,\mathrm{in,SWB}}$ and $\sigma_{n,\mathrm{out,SWB}}$ determine the widths of the interior shoulder and the exterior shock jump, respectively.

The magnetic structure of the bubble is governed by flux freezing. As the wind bubble inflates, the frozen-in magnetic field lines are stretched and diluted within the volume, while being strongly compressed into the dense shell at the boundary. We model this as:

\begin{multline}
B_{0,\mathrm{SWB}}(R,Z) = 
\mathcal{G}(d_{\mathrm{SWB}}; A_{B,\mathrm{cav,SWB}}, \sigma_{B,\mathrm{cav,SWB}}) \\
+ \mathcal{S}(d_{\mathrm{SWB}}; A_{B,\mathrm{sh,SWB}}, R_{\mathrm{sh,SWB}}, \sigma_{B,\mathrm{in,SWB}}, \sigma_{B,\mathrm{out,SWB}}).
\end{multline}
Here, the Gaussian term with negative amplitude $A_{B,\mathrm{cav,SWB}}$ accounts for the magnetic depletion in the cavity. For the shell component, $A_{B,\mathrm{sh,SWB}}$ denotes the peak magnetic field strength at the shell radius $R_{\mathrm{sh,SWB}}$, while $\sigma_{B,\mathrm{in,SWB}}$ and $\sigma_{B,\mathrm{out,SWB}}$ determine the widths of the interior and exterior magnetic gradients, respectively.

Standard SWB models describe the interior as a region of hot, shocked stellar wind ($T \sim 10^6$ K) that maintains high pressure to support the bubble \citep{WEAVERInterstellar1977}. Because the sound speed is high, the interior evolves into an isobaric region with a nearly uniform temperature distribution. To reproduce this characteristic flat-top thermal profile, we model the interior temperature using the super-Gaussian plateau function defined in Eq. \ref{eq:generic_supergauss} in the main text.
The total electron temperature at SWB includes this plateau and the shock-heated shell:
\begin{multline}
T_{e,\mathrm{SWB}}(R,Z) = \mathcal{P}(d_{\mathrm{SWB}}; A_{T,\mathrm{plat,SWB}}, R_{\mathrm{flat,SWB}}, s) \\
+ \mathcal{S}(d_{\mathrm{SWB}}; A_{T,\mathrm{sh,SWB}}, R_{\mathrm{sh,SWB}}, \sigma_{T,\mathrm{in,SWB}}, \sigma_{T,\mathrm{out,SWB}}),
\end{multline}
where $A_{T,\mathrm{plat,SWB}}$ is the temperature amplitude of the hot bubble and $R_{\mathrm{flat,SWB}}\, (\approx 0.6\, R_{\mathrm{sh,SWB}})$ determines the extent of the isothermal region (using index $s=8$). For the shell component, $A_{T,\mathrm{sh,SWB}}$ denotes the temperature excess at the shock front, while $\sigma_{T,\mathrm{in,SWB}}$ and $\sigma_{T,\mathrm{out,SWB}}$ characterize the thermal gradients on the interior and exterior sides, respectively.

\begin{table*}[!h]
\centering
\caption{Adopted background (WIM) parameters and positron-fraction parameters.}
\label{tab:param_background}

\begin{tabular}{lll}
\hline
\textbf{WIM Parameters} & \textbf{Value} & \textbf{Meaning / units} \\
\hline
$R_\odot$              & 8.5   & Solar circle radius (kpc) \\
$n_1$                  & 0.03  & Thin-disk density (cm$^{-3}$) \\
$h_1$                  & 0.1   & Thin-disk scale height (kpc) \\
$n_2$                  & 0.1   & Thick-disk/WIM density (cm$^{-3}$) \\
$h_2$                  & 1.0   & Thick-disk/WIM scale height (kpc) \\
$L_2$                  & 15.0  & Radial scale length of WIM density (kpc) \\
$B_\odot$              & 5.0   & Field at $R=R_\odot, Z=0$ ($\mu$G) \\
$L_B$                  & 10.0  & Radial scale length of disk field (kpc) \\
$h_{B,1}$              & 1.0   & Vertical scale height of disk field (kpc) \\
$B_{\text{halo}}$      & 2.0   & Base halo field ($\mu$G) \\
$h_{B,2}$              & 3.0   & Vertical scale height of halo field (kpc) \\
$T_{\text{e,bkg}}$     & 8000  & Background temperature (K) \\
$p_{\text{bkg}}$       & 0.01  & Background positron fraction \\
$p_{\text{peak}}$      & 0.2   & Peak positron fraction at source \\
$\sigma_{p,\text{scale}}$ & 1.0 & Galactic positron source scale (kpc) \\
\hline
\end{tabular}
\end{table*}

\begin{table*}[!h]
\caption{Summary of analytical profile parameters for embedded astrophysical structures and corresponding numerical values used in 
Figs. \ref{fig:B_ni_Te_beta_3D}--\ref{fig:SNR_1d_beta_amp_width}. 
For each structure and physical quantity, the table lists the analytical form used and the specific amplitude and scale parameters. Unless otherwise noted, lengths are in kpc, number densities are in cm$^{-3}$, magnetic field strengths are in $\mu$G, and temperatures are in K. Note that amplitudes $A<0$ represent depletion. The central coordinates are $(R_1, Z_1)$ for H\,II regions, $(R_2, Z_2)$ for SWBs, and $(R_2, Z_2)$ for SNRs.}
\label{tab:profiles_parameters}
\centering
\begin{tabular}{llll}
\hline
\hline
Structure & Quantity & Analytical Form & Parameters \\
\hline
H\,II Region
& Electron density $n_{e,\mathrm{HII}}$
& $\mathcal{G}$
& $A_{n,\mathrm{HII}}=\text{2000 $\times$ $n_{e,\mathrm{bkg}}(R_1,Z_1)$}$, $\sigma_{n,\mathrm{HII}}=0.25$. \\
& Magnetic field $B_{0,\mathrm{HII}}$
& $\mathcal{G}$
& $A_{B,\mathrm{HII}}=\texttt{$-0.7\times B_{0,\mathrm{bkg}}(R_1,Z_1)$}$, $\sigma_{B,\mathrm{HII}}=0.25$. \\
& Temperature $T_{\mathrm{HII}}$
& $\mathcal{G}$
& $A_{T,\mathrm{HII}}= 0.25\,\times\,T_{e,\text{bkg}}$, $\sigma_{T,\mathrm{HII}}=0.25$. \\
\hline
\cline{2-4}
SWB Shell
& Electron density $n_{e,\mathrm{sh,SWB}}$
& $\mathcal{S}$
& $A_{n,\mathrm{sh,SWB}}=\texttt{$4\times n_{e,\mathrm{bkg}}(R_2,Z_2)$}$,
  $R_{\mathrm{sh,SWB}}=\texttt{0.25}$,
\\
&     &  & $\sigma_{n,\mathrm{in,SWB}}=\texttt{$R_{\mathrm{sh,SWB}}/10$}$,
$\sigma_{n,\mathrm{out,SWB}}=\texttt{$\sigma_{n,\mathrm{in,SWB}}/4$}$.  \\
& Magnetic field $B_{0,\mathrm{sh,SWB}}$
& $\mathcal{S}$
& $A_{B,\mathrm{sh,SWB}}=\texttt{$10\times B_{0,\mathrm{bkg}}(R_3,Z_3)$}$,
 \\
&     &     & $\sigma_{B,\mathrm{in,SWB}}=\texttt{$R_{\mathrm{sh,SWB}}/10$}$,
$\sigma_{B,\mathrm{out,SWB}}=\texttt{$\sigma_{B,\mathrm{in,SWB}}/4$}$.  \\
& Temperature $T_{e,\mathrm{sh,SWB}}$
& $\mathcal{S}$
& $A_{T,\mathrm{sh,SWB}}=\texttt{$10^4\,$K$-T_{e,\mathrm{bkg}}$}$,
\\
&     &    &    $\sigma_{T,\mathrm{in,SWB}}=\texttt{$R_{\mathrm{sh,SWB}}/2$}$, $\sigma_{T,\mathrm{out,SWB}}=\texttt{$\sigma_{n,\mathrm{in,SWB}}/6$}$.  \\
\cline{2-4}
SWB Cavity
& Electron density $n_{e,\mathrm{cav,SWB}}$
& $\mathcal{G}$
& $A_{n,\mathrm{cav,SWB}}=\texttt{$-0.9\times n_{e,\mathrm{bkg}}(R_2,Z_2)$}$,
\\
&    &    & 
$\sigma_{n,\mathrm{cav,SWB}}=\texttt{$R_{\mathrm{sh,SWB}}/2$}$. \\
& Magnetic field $B_{0,\mathrm{cav,SWB}}$
& $\mathcal{G}$
& $A_{B,\mathrm{cav,SWB}}=\texttt{$-0.5\times B_{0,\mathrm{bkg}}(R_2,Z_2)$}$,\\
&  &   & $\sigma_{B,\mathrm{cav,SWB}}=\texttt{$R_{\mathrm{sh,SWB}}/2$}$.   \\
\cline{2-4}
SWB Interior
& Temperature $T_{e,\mathrm{plat,SWB}}$
& $\mathcal{P}$
& $A_{T,\mathrm{plat,SWB}}=\texttt{$10^6\,$K$-T_{e,\mathrm{bkg}}$}$,
  $R_{\mathrm{flat,SWB}}=\texttt{$0.6\times R_{\mathrm{sh,SWB}}$}$.
\\
\hline
SNR Shell
& Electron density $n_{e,\mathrm{sh,SNR}}$
& $\mathcal{S}$
& $A_{n,\mathrm{sh,SNR}}=\texttt{$4\times n_{e,\mathrm{bkg}}(R_3,Z_3)$}$,
$R_{\mathrm{sh,SNR}}=0.25$,\\
&     & &$\sigma_{n,\mathrm{in,SNR}}=\texttt{$R_{\mathrm{sh,SNR}}/10$}$,
$\sigma_{n,\mathrm{out,SNR}}=\texttt{$\sigma_{n,\mathrm{in,SNR}}/10$}$. \\
& Magnetic field $B_{0,\mathrm{sh,SNR}}$
& $\mathcal{S}$
& $A_{B,\mathrm{sh,SNR}}=\texttt{$4\times B_{0,\mathrm{bkg}}(R_3,Z_3)$}$,
\\
&    &   & $\sigma_{B,\mathrm{in,SNR}}=\texttt{$R_{\mathrm{sh,SNR}}/10$}$, $\sigma_{B,\mathrm{out,SNR}}=\texttt{$\sigma_{B,\mathrm{in,SNR}}/10$}$.\\
& Temperature $T_{e,\mathrm{sh,SNR}}$
& $\mathcal{S}$
& $A_{T,\mathrm{sh,SNR}}=\texttt{$(10^4\,$K$-T_{e,\mathrm{bkg}})$}$,
\\
&    &       & $\sigma_{T,\mathrm{in,SNR}}=\texttt{$R_{\mathrm{sh,SNR}}/1.5$}$,
$\sigma_{T,\mathrm{out,SNR}}=\texttt{$\sigma_{T,\mathrm{in,SNR}}/10$}$. \\
\cline{2-4}
SNR Cavity
& Electron density $n_{e,\mathrm{cav,SNR}}$
& $\mathcal{G}$
& $A_{n,\mathrm{cav,SNR}}=\texttt{$-\,n_{e,\mathrm{bkg}}(R_3,Z_3)$}$,
$\sigma_{n,\mathrm{cav,SNR}}=\texttt{$R_{\mathrm{sh,SNR}}/3$}$. \\
\cline{2-4}
SNR PWN
& Electron density $n_{e,\mathrm{PWN,SNR}}$
& $\mathcal{G}$
& $A_{n,\mathrm{PWN,SNR}}=\texttt{$n_{e,\mathrm{bkg}}(R_3,Z_3)/1.25$}$,
 \\
&   &  &$\sigma_{n,\mathrm{PWN,SNR}}=\texttt{$R_{\mathrm{sh,SNR}}/10$}$.   \\
& Magnetic field $B_{0,\mathrm{PWN,SNR}}$
& $\mathcal{G}$
& $A_{B,\mathrm{PWN,SNR}}=\texttt{120}\,\mu\mathrm{G}$,
$\sigma_{B,\mathrm{PWN,SNR}}=\texttt{$R_{\mathrm{sh,SNR}}/25$}$. \\
& Positron fraction $p_{\mathrm{SNR}}$
& $\mathcal{G}$
& $A_{p,\mathrm{SNR}}=\texttt{$(p_{\mathrm{peak,SNR}}-p_{\mathrm{bkg}})$}$,  \\
&    &    &  with $\texttt{$p_{\mathrm{peak,SNR}}=10^4$}$,
$\sigma_{p,\mathrm{SNR}}=\texttt{$R_{\mathrm{sh,SNR}}/25$}$.   \\
& Temperature $T_{e,\mathrm{core,SNR}}$
& $\mathcal{G}$
& $A_{T,\mathrm{core,SNR}}=\texttt{$(10^7+T_{e,\mathrm{bkg}})$}$,
$\sigma_{T,\mathrm{core,SNR}}=\texttt{$R_{\mathrm{sh,SNR}}/5$}$. \\
\hline
\hline
\end{tabular}
\end{table*}

\newpage

\section{First Order Equations: \label{App.-A}}

\begin{equation}\label{eq16}
-a_{1} \lambda \frac{\partial \psi^{(1)}}{\partial \xi}+l_{x} \frac{\partial v_{ix}^{(1)}}{\partial \xi}+l_{z} \frac{\partial v_{iz}^{(1)}}{\partial \xi}=0,
\end{equation}
\begin{equation}\label{eq17}
v_{ix}^{(1)}= \beta \lambda l_{x} \frac{\partial^{2} \phi^{(1)}}{\partial \xi^{2}},
\end{equation}
\begin{equation}\label{eq18}
\lambda \frac{\partial v_{iz}^{(1)}}{\partial \xi}= \beta l_{z} \frac{\partial \psi^{(1)}}{\partial \xi},
\end{equation}
\begin{equation}\label{eq19}
\beta l_{x}^{(2)} l_{z}^{(2)} \frac{\partial^{2} \phi^{(1)}}{\partial \xi^{2}}= a_{1} \lambda^{2} \psi^{(1)}- l_{z} \lambda v_{iz}^{(1)}.
\end{equation}

\section{Second Order Equations: \label{App.-B}}

\begin{eqnarray}\label{eq23}
-a_{1} \lambda \frac{\partial \psi^{(2)}}{\partial \xi}-2a_{2} \lambda \psi^{(1)} \frac{\partial \psi^{(1)}}{\partial \xi}+a_{1} \frac{\partial \psi^{(1)}}{\partial \tau}
+l_{x} \frac{\partial v_{ix}^{(2)}}{\partial \xi}\\ \nonumber
+a_{1}l_{x} \frac{\partial (\psi^{(1)}v_{ix}^{(1)})}{\partial \xi}+l_{z} \frac{\partial v_{iz}^{(2)}}{\partial \xi}
+a_{1}l_{z} \frac{\partial (\psi^{(1)}v_{iz}^{(1)})}{\partial \xi}=0
\end{eqnarray}
\begin{equation}\label{eq24}
v_{ix}^{(2)}= \beta \left(\lambda l_{x} \frac{\partial^{2} \phi^{(2)}}{\partial \xi^{2}}-l_{x}\frac{\partial^{2} \phi^{(1)}}{\partial \tau \partial \xi}\right)
\end{equation}
\begin{equation}\label{eq25}
-\lambda \frac{\partial v_{iz}^{(2)}}{\partial \xi}+\frac{\partial v_{iz}^{(1)}}{\partial \tau}+l_{x}v_{ix}^{(1)}\frac{\partial v_{iz}^{(1)}}{\partial \xi}+l_{z}v_{iz}^{(1)}\frac{\partial v_{iz}^{(1)}}{\partial \xi}= -\beta l_{z} \frac{\partial^{2} \psi^{(2)}}{\partial \xi^{2}}
\end{equation}
\begin{eqnarray}\label{eq26}
\beta l_{x}^{(2)} l_{z}^{(2)} \frac{\partial^{4} (\phi^{(2)}-\psi^{(1)})}{\partial \xi^{4}}=a_{1} \lambda^{2}\frac{\partial^{2}\psi^{(2)}}{\partial \xi^{2}}+a_{2} \lambda^{2}\frac{\partial^{2}\psi^{(1)}}{\partial \xi^{2}}\\ \nonumber
-2a_{1}\lambda\frac{\partial^{2}\psi^{(1)}}{\partial \xi \partial \tau}-l_{z}\lambda \frac{\partial^{2} v_{iz}^{(2)}}{\partial \xi^{2}}+l_{z}\frac{\partial^{2} v_{iz}^{(1)}}{\partial \xi \partial \tau}-a_{1}l_{z}\lambda\frac{\partial^{2}(v_{iz}^{(1)}\psi^{(1)})}{\partial \xi^{2}}
\end{eqnarray}

\end{document}